\documentclass[twocolumn]{article}

\usepackage{fullpage}
\usepackage{graphicx}% Include figure files
\usepackage{amsmath}
\usepackage{amssymb}
\usepackage[utf8]{inputenc}
\usepackage{authblk}
\usepackage{abstract}

\newcommand{\im}{\mathrm{Im}}
\newcommand{\re}{\mathrm{Re}}

\title{Laplacian networks: growth, local symmetry and shape optimization}% Force line breaks with \\

\author[1]{O.~Devauchelle}
\affil[1]{Institut de Physique du Globe de Paris, France}

\author[2]{P.~Szymczak}
\affil[2]{Institute of Theoretical Physics, Faculty of Physics, University of Warsaw, Poland}
\author[2]{M.~Pecelerowicz}

\author[3]{Y.~Cohen}
\affil[3]{Lorenz Center, Department of Earth, Atmospheric, and Planetary Sciences, Massachusetts Institute of Technology, Cambridge, MA, USA}
\author[4]{H.J.~Seybold}
\affil[4]{Institute of Terrestrial Ecosystems, Department of Environmental Systems Science, ETH Zurich, Switzerland}
\author[3]{D.H.~Rothman}

\date{\today}

\begin{document}

\twocolumn[

\maketitle

\begin{onecolabstract}
Inspired by river networks and other structures formed by Laplacian growth, we use the Loewner equation to investigate the growth of a network of thin fingers in a diffusion field. We first review previous contributions to illustrate how this formalism reduces the network's expansion to three rules, which respectively govern the velocity, the direction, and the nucleation of its growing branches. This framework allows us to establish the mathematical equivalence between three formulations of the direction rule, namely geodesic growth, growth that maintains local symmetry and growth that maximizes flux into tips for a given amount of growth. Surprisingly, we find that this growth rule may result in a network different from the static configuration that optimizes flux into tips.\\
\end{onecolabstract}
]

\section{Introduction}

When an interface moves in response to a quantity diffusing towards it, the coupling of its deformation with the flux that moves it generates remarkable dynamics. The Saffman-Taylor experiment is a paragon of this class of systems \cite{saffman1958penetration}, but many other occurrences manifest themselves in nature through beautifully ramified patterns, such as wormholes in dissolving rocks \cite{mcduff2010,szymczak2013interacting}, finger-like smoldering in combustion experiments~\cite{Zik:1999}, metallic dendrites grown by electrochemical deposition~\cite{Brady1984,Matsushita1984}, the formation of lungs \cite{lubkin1995mechanism,clement2012branching} or the growth of root systems \cite{fitter1987architectural,futsaether2002growth,hodge2009plant}. A similar process takes place at a much larger scale when a river drains the groundwater that surrounds it \cite{dunne1980formation,abrams2009growth,devauchelle2012ramification,petroff2013bifurcation}. The seeping water can erode sediments away from springs, thus causing the stream to grow further, to bifurcate, and ultimately to carve a dendritic drainage network into the surrounding landscape.

Even when the velocity of the boundary depends linearly on the diffusion flux, the deformation of the domain makes systems of this class non-linear. This non-linearity can generate intricate dendrites growing in competition with each other. The emergence of multiple scales and collective behaviors from an extremely simple set of equations has attracted ongoing interest \cite{arneodo1989uncovering, derrida1992needle, hastings2001growth,Mathiesen2006}.

The dendrites generated by diffusion-limited growth result from the concentration of the flux at the tip of a finger, much like the electric field around a lightning rod. As a consequence, most of the growth occurs at the tips. The utmost limit of such systems is a diffusion field drained by a network of infinitely thin fingers, whose tips concentrate the diffusion field into a singularity \cite{Kurtze:1991,Peterson:1998,Peterson:1999,carleson2002laplacian,gubiec2008fingered}. At the cost of this singularity however, we simplify two aspects of our problem. First, we eliminate the ultraviolet catastrophe caused by the Saffman-Taylor instability, which generates cusps from virtually every initial configuration \cite{Bensimon1986,dawson1994long,Bauer2003}.
%In contrast, a classical regularization, such as surface tension, renders the mathematical formulation dependent on specific physics and adds a new layer of complexity to the initial problem \cite{combescot1988analytic}.
Second, the singular harmonic growth reduces the motion of a boundary to the trajectories of a denumerable number of tips, making it easier to grasp the dynamics of the system and to understand their physical implications.

To grow a network of infinitesimally thin fingers, we need to specify (i) the velocity of each tip, (ii) the direction of its growth and (iii) when it branches. These rules control the dynamics of the network, and consequently mark its geometry. For instance, a finger growing along the flow lines of a diffusion field bifurcates at an angle of $2\pi/5$ \cite{hastings2001growth,carleson2002laplacian}. River networks cut by seepage erosion exhibit this property unambiguously \cite{devauchelle2012ramification,petroff2013bifurcation}, thus illustrating how the growth dynamics of a network can be inferred from its final shape.

By growing along a flow line, a finger balances the fluxes coming from both sides of its tip. In fact, these two propositions are equivalent. For instance, in a wording inspired from fracture mechanics, we can say that rivers formed by seepage erosion grow according to the principle of local symmetry (PLS) \cite{barenblatt1961brittle,yossi}. This reformulation explicitly constrains the geometry of a network during its growth, and can therefore be used to identify the velocity and bifurcation rules based on its shape, at least in principle.

The analogy with fracture mechanics also suggests an alternative interpretation of the PLS: some cracks are hypothesized to propagate in the direction that maximizes the release of elastic energy \cite{erdogan1963crack,adda2005brittle}. Do river networks or dissolution wormholes satisfy a similar optimization principle? To investigate this question, we use the formalism introduced by Loewner to describe the growth of fingers in Laplacian fields \cite{lowner1923untersuchungen}. Indeed, in many problems, the field driving the growth satisfies Laplace's equation, making complex analysis a convenient framework to describe the dynamics of two-dimensional analogues of the Saffman-Taylor experiment. In particular, an attractive idea is to represent the deformation of the domain by a conformal map which, by construction, satisfies Laplace's equation \cite{lowner1923untersuchungen,gruzberg2004loewner,Gustafsson2006}. The so-called ``Loewner equation'' describes the evolution of the conformal map; its expression for a specific system depends on how its boundary grows \cite{Loewner1923, kufarev1943, gustafsson2014}.

Here, we first introduce the Loewner equation and review some of its fundamental properties, focusing on physical and geometrical interpretation, rather than on rigorous mathematical derivation, referring the reader inclined to the latter to appropriate references (Section~\ref{sec:loewner_growth}). We then use this formalism to connect the dynamics of growing fingers with the resulting geometry of the network they compose, with special attention to the rule controlling the direction of their growth (Section~\ref{sec:growth_rules}). Finally, we show that, in general, the dynamical optimization of flux does not result in an optimal static network (Section~\ref{sec:opti_branch}).

\section{Loewner growth}\label{sec:loewner_growth}

This section provides a review of growth processes described by the Loewner equation. By presenting simple examples from the
literature \cite{kager2004exact, derrida1992needle, gubiec2008fingered}, we describe growth as a consequence of three ingredients: the growth factors of individual
tips or fingers, the direction in which this growth occurs, and the ramification of tree-like networks.
We begin with a brief review of the Loewner equation, and use it to represent the growth of a single finger.  We show how growth depends on the motion, or lack thereof, of the image of the finger's tip on the real axis of the mathematical plane.  We then proceed through examples of a single finger's growth, to, ultimately, the interaction of multiple fingers.
Readers who are already familiar with the literature on the deterministic Loewner
equation  may find it efficient to skip ahead to section~\ref{sec:growth_rules} \cite{carleson2002laplacian, kager2004exact, gubiec2008fingered}.

\subsection{The Loewner equation}

\begin{figure}
\centering
\includegraphics[width = \linewidth]{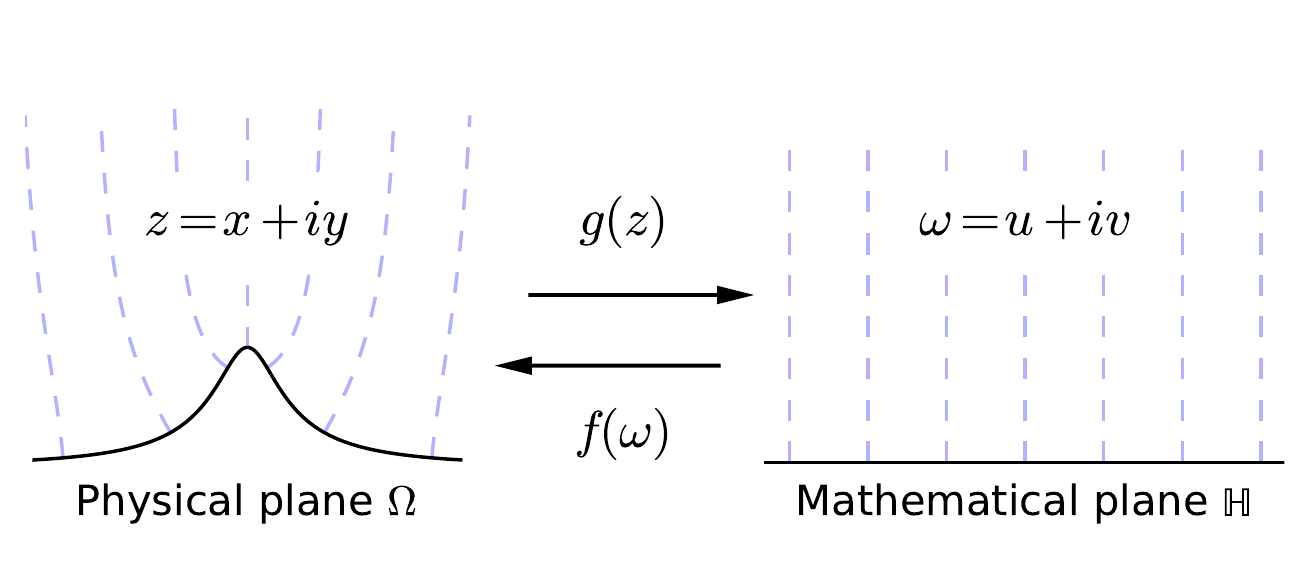}
\caption{Conformal mapping from the physical plane (left) to the upper half plane (right). Black solid line: absorbing boundary, blue dashed lines: streamlines for the diffusive field.  Coordinates in the physical (resp. mathematical) plane are represented by the complex number $z$ (resp. $\omega$).\label{fig:general_mapping}}
\end{figure}

In the Saffman-Taylor experiment, a viscous liquid is confined between two parallel glass plates separated by a gap of uniform thickness \cite{saffman1958penetration}. Air, injected at constant pressure, pushes the liquid, which flows at a velocity proportional to the pressure gradient. Mass balance then requires the pressure field to be harmonic in the liquid, and the air-fluid interface to move at the fluid velocity.

We may formalize this problem as follows. A smooth interface bounds a domain $\Omega$ where the field $p$ is harmonic, and deforms in response to the flux $\nabla p$ that reaches it. The deformation of the boundary thus depends on its own shape through Laplace's equation.

In two dimensions, complex analysis combines the two facets of Laplacian growth, namely the motion of the boundary and the evolution of the diffusion field. Indeed, the Riemann mapping theorem tells us that there exist an analytical function $g$ that maps the domain $\Omega$ (the physical plane, figure~\ref{fig:general_mapping}) onto the upper half of the complex plane $\mathbb{H}$ (the mathematical plane). If we further require that the far field behave like $z$, that is
\begin{equation}
\lim_{z \rightarrow \infty}(g(z)-z)=0 \, ,
\end{equation}
then the mapping $g$ is unique. By construction, the imaginary part of $g$ is harmonic and can serve as the Laplacian field $p$ (if the interface is absorbing). Likewise, its real part is the stream function associated to $p$. The mapping $g$ thus encodes both the shape of the domain and the diffusive field.

In some cases, we can express the change of $g$ over time as an explicit evolution equation due to Loewner  and Kufarev \cite{Loewner1923,kufarev1943} (Gustafsson discusses their respective contributions \cite{gustafsson2014}). To illustrate this, we first define $f$ as the inverse mapping of $g$ (figure~\ref{fig:general_mapping}). The function $f$ maps any horizontal line in the mathematical plane onto an equipotential line in the physical plane. In particular, it maps the real axis onto the growing interface. As a consequence, its derivative $f'$ aligns with the interface, and $f'/\left| f'\right|$ is the unit complex number tangent to it. Based on this observation, we express the normal velocity $v_n$ of the growing interface as the scalar product of $\dot{f}$ with this unit number:
\begin{equation}
v_n = - \im \left( \frac{\dot{f}}{f'} \right) \left| f' \right| \, .
\end{equation}
We then define the growth function $\varphi$, analytical in the upper half plane, such that
\begin{equation}
\im \left( \frac{1}{\varphi } \right) =  \dfrac{v_n} {\left| f' \right|}
\label{eq:phi_vel}
\end{equation}
on the real axis. Finally, the growth equation reads
\begin{equation}
\dot{f} = -\dfrac{f'}{\varphi}
\label{eq:fpfd}
\end{equation}
where the growth function $\varphi$ encapsulates both the interface motion and the far-field boundary conditions. If, for instance, the interface moves in proportion to the diffusive flux ($v_n \propto |\nabla p |$), $f$ satisfies the so-called Polubarinova-Galin \cite{polubarinova1945,galin1945} (or Laplacian growth) equation, which represents the simplest possible model of the Saffman-Taylor experiment. Most of its solutions become singular at finite time, although a special class of analytical solutions grow continuous, finite-width fingers \cite{dawson1994long}.

Here, we derive the Loewner equation describing the growth of a collection of infinitely thin fingers as a special case of equation (\ref{eq:fpfd}), where growth concentrates at the fingers' tips \cite{carleson2002laplacian,gruzberg2004loewner,gubiec2008fingered}. In this context, equation (\ref{eq:fpfd}) is referred to as the Loewner equation \cite{lowner1923untersuchungen}. It is usually expressed in terms of the forward mapping $g$ instead of its inverse $f$, but the two formulations are equivalent. Indeed, taking the time derivative of $f(g(z))=z$, we find that $\dot{g}=-\dot{f}/f'$, and we can formulate the Loewner equation in a more familiar way \cite{gubiec2008fingered}:
\begin{equation}
\dot{g}=\dfrac{1}{\varphi \circ g } \, .
\label{eq:loewner_inverse}
\end{equation}

\subsection{Growth of a finger}\label{sec:growth}

\begin{figure}
\centering
\includegraphics[width = \linewidth]{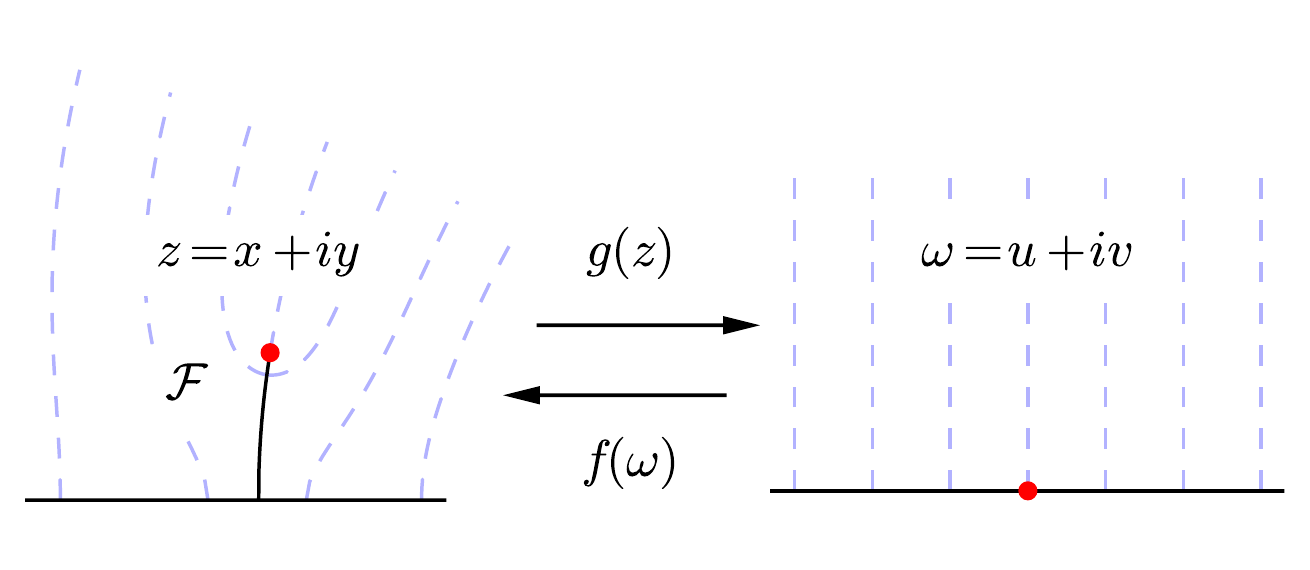}
\caption{Mapping of a growing finger $\cal N$ onto the real axis by $g$ and its inverse $f$. Red dots indicate the figer's tip $\gamma$ (left) and the corresponding pole $a$ in the mathematical plane (right).\label{fig:def_f}}
\end{figure}

Let us first consider a single finger $\mathcal{F}$ growing off of the real axis, into the upper part of the complex plane (figure~\ref{fig:def_f}). This finger is a curved segment connecting the real axis to its tip $\gamma$. As it grows, the path followed by $\gamma$ through time sets its shape.

The analytical function $g$ maps the region outside the finger $\Omega = \mathbb{H}\backslash \mathcal{F} $ onto the upper half of the complex plane $\mathbb{H}$:
\begin{equation} 
g:\mathbb{H}\backslash \mathcal{F} \rightarrow \mathbb{H}
\end{equation}
Geometrically, this conformal map cuts the finger open and projects its sides onto the real axis.

The mapping $g$ is essentially unique if we impose appropriate boundary conditions. Interpreting the imaginary part of $g$ as the harmonic field, we require it to vanish along the finger and the real axis. In addition, the condition
\begin{equation}
g(z) \rightarrow  z + O(1/z), \quad z \rightarrow \infty
\label{eq:far_field_BC}
\end{equation}
corresponds to a constant diffusion flux far away from the boundary. It also fixes the constant inherent to the definition of the stream function. We use this boundary condition hereafter.

As the finger grows, the mapping $g$ evolves. It is this evolution that we wish to describe with the Loewner equation. For convenience, we will derive the Loewner equation of its inverse $f$. This analytical function can be continuously extended along the real axis. It is then two-to-one along the finger's sides, and one-to-one at its tip. The tip can therefore be added to the domain of $g$; its corresponding image $a=g(\gamma)$ is called ``pole'' (figure~\ref{fig:def_f}). By construction, the pole $a$ is a real number, and
\begin{equation}
f(a)=\gamma \, .
\label{eq:defa}
\end{equation}

If a point slides along the real axis in the mathematical plane, its image follows the finger's side in the physical plane. As it passes the pole $a$, its image makes a sharp U-turn around the finger's tip. This implies that the derivative of $f$  vanishes at the pole:
\begin{equation}
f'(\omega) = f''(a) (\omega-a) + {\cal O} (\omega-a)^2 \, .
\label{eq:tipdef}
\end{equation}
Differentiating equation (\ref{eq:defa}) with respect to time, and assuming temporarily that the pole and its image move continuously, we find that the time derivative of the inverse mapping sets the velocity of the finger's tip:
\begin{equation}
\dot{\gamma} = \dot{f}(a) \,.
\label{eq:gamma_dot}
\end{equation}
In this expression, the pole $a$ is a function of time.

We now require that the boundary deforms at the finger's tip only. This means that the normal velocity $v_n$ vanishes everywhere, except possibly at the finger's tip. After equation (\ref{eq:phi_vel}), this condition is more conveniently formulated by requiring that
\begin{equation}
\im \left( \dfrac{1}{\varphi} \right) = 0
\label{eq:bc_phi}
\end{equation}
on the real axis but not at the pole $a$. More specifically, we want the normal velocity to be finite at the pole. After equations (\ref{eq:phi_vel}) and (\ref{eq:tipdef}), this can be true only if $a$
is a pole of the growth function:
\begin{equation}
\dfrac{1}{\varphi} = \dfrac{G}{\omega - a} + {\cal O} (1)
\label{eq:phi_near_pole}
\end{equation}
where $G$ is a positive real number. The normal velocity of the growing boundary then vanishes everywhere, except at the finger's tip, where it is $v_n = \left| G f''(a) \right|$. The growth velocity being proportional to $G$, one usually refers to the latter as the ``growth factor''. This accords with equation (\ref{eq:gamma_dot}), from which
\begin{equation}
\dot{\gamma} = G f''(a) \, .
\label{eq:gamma_dot_G}
\end{equation}
Since $G$ is real, the finger grows in the direction of its tip at first order. Section \ref{sec:growth_rules} is devoted to understanding how the finger bends out of this alignment at the next order.

Considering momentarily a single growing finger, $a$ must be the only zero of $\varphi$ in $\mathbb{H} \cup \mathbb{R}_{+}$. Further requiring that the growth function preserves the far-field boundary condition (\ref{eq:far_field_BC}), we find that $\varphi$ is exactly
\begin{equation}
\dfrac{1}{\varphi} = \dfrac{G}{\omega - a} \, .
\end{equation}
We next illustrate how the Loewner equation works, using simple expressions for the growth function $\varphi$.

\subsection{Static pole}\label{sec:static_pole}

In general, the growth factor $G$ and the pole $a$ are functions of time. However, let us first keep them constant, and assume that $f$ is initially the identity mapping. Then, the Loewner equation has a simple solution:
\begin{equation}
f= a + \sqrt{ (\omega-a)^2 - 2 G t }
\label{eq:strait_needle}
\end{equation}
where the branch of the square root is such that $ f(\omega) \sim \omega $ for large $\omega$. This solution corresponds to the growth of a straight finger off of the real axis while remaining orthogonal to it, at velocity $\sqrt{G/(2t)}$. The nascent tip grows infinitely fast initially ($t=0$), before it is slowed down by the changes it induces on the diffusion field.

\subsection{Linear forcing}

\begin{figure}
\centering
\includegraphics[width = \linewidth]{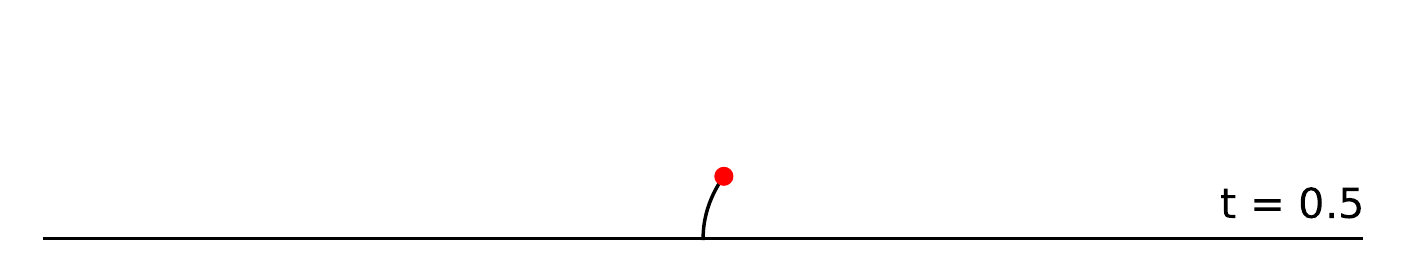}\\
\medskip
\includegraphics[width = \linewidth]{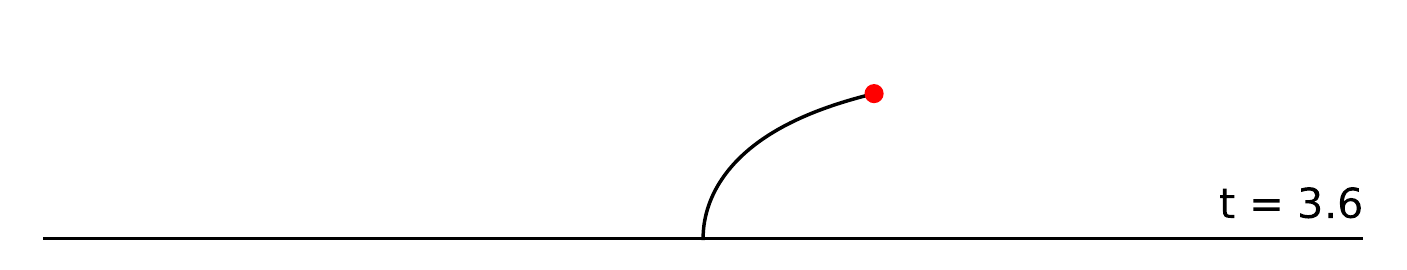}\\
\medskip
\includegraphics[width = \linewidth]{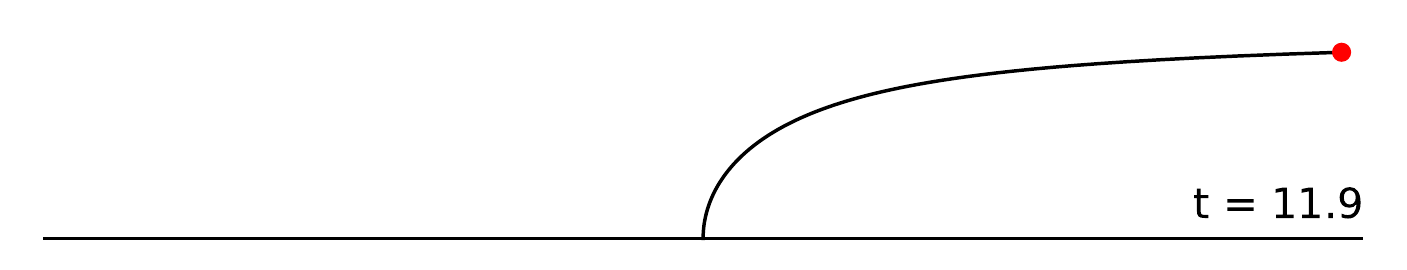}\\
\caption{Loewner growth of a finger, with linear forcing, corresponding to equation (\ref{eq:gamma_linear_forcing}).\label{fig:loewner_linear_forcing}}
\end{figure}

As long as we consider a single finger, the growth factor can be included in a redefinition of time, as can be readily checked with the Loewner equation \cite{gubiec2008fingered}. Physically, this observation means that the growth velocity of an isolated finger has no influence on its shape.

The motion of the pole, on the contrary, controls the shape of the growing finger ---hence its alternative name, ``driving function''. For instance, we can make the pole move at constant velocity along the real axis, while keeping $G$ constant:
\begin{equation}
a=t,
\quad
G = 1 \, .
\end{equation}
This system is still analytically tractable \cite{kager2004exact}. The solution to the corresponding Loewner equation, if $f$ is initially the identity, reads
\begin{equation}
f=h^{-1} \left( h(\omega - t) - t \right)
\label{eq:first_composition}
\end{equation}
where
\begin{equation}
h(\omega) = - \omega - \log (1 - \omega) \, .
\end{equation}
These equations lead to an implicit expression for the trajectory of the finger in the physical space:
\begin{equation}
\gamma + \log (1 - \gamma) = t \, .
\label{eq:gamma_linear_forcing}
\end{equation}
As it grows, the finger is bent by the motion of the pole $a$ in the mathematical plane (figure~\ref{fig:loewner_linear_forcing}). The motion of the pole therefore pilots the growth of the finger \cite{kager2004exact,gruzberg2004loewner}.

\subsection{Discontinuous forcing}\label{sec:discondinuous}

\begin{figure}
\centering
\includegraphics[width = 0.9\linewidth]{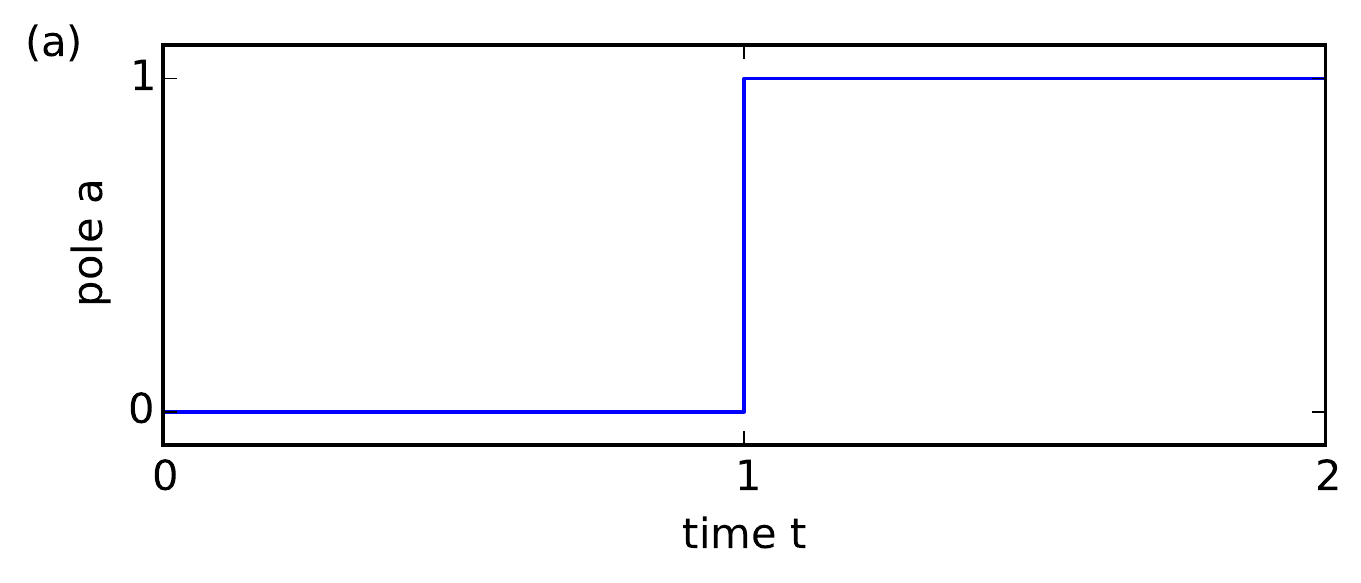}\\
\includegraphics[width = 0.32\linewidth]{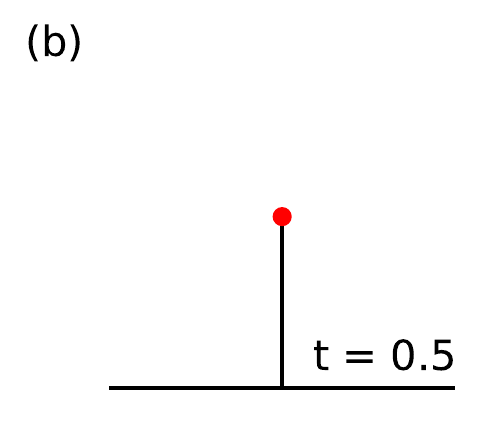}
\includegraphics[width = 0.32\linewidth]{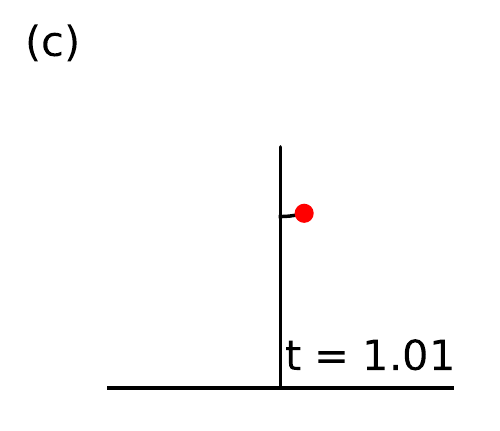}
\includegraphics[width = 0.32\linewidth]{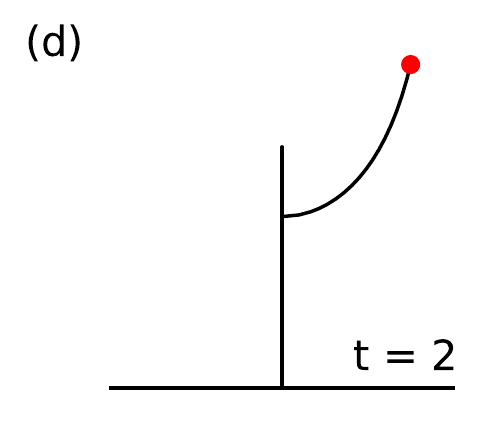}
\caption{Loewner growth of a finger with discontinuous forcing, according to equation (\ref{eq:discontinuous_loewner}). The forcing (a) first generates a single straight finger (b). The discontinuity occurring at time $t=1$ nucleates a second branch, while the first branch becomes inactive (c,d).\label{fig:discontinuous_growth}}
\end{figure}

We now turn our attention to the discontinuous motion of a pole, of which the step function is a simple example (figure~\ref{fig:discontinuous_growth}):
\begin{equation}
a_d = \left\{
\begin{array}{ccc}
0 & \mbox{if} & 0 \leq t < t_c \\ 
a_0 & \mbox{if} & t \geq t_c
\end{array} 
\right. \, .
\end{equation}
Again, this problem can be solved analytically \cite{kager2004exact}, assuming $G=1$. To do so, we first introduce the invariance of the Loewner evolution with respect to composition \cite{gruzberg2004loewner}. If $f_t$ is a solution of the Loewner equation, and $f_*$ an analytical function independent of time, the composed function $f_* \circ f_t$ is another solution, only with different initial conditions. This property allows us to transform any solution of the Loewner equation into another solution with a different initial state. Equation (\ref{eq:first_composition}) actually results from this procedure, which we use again to construct the solution to the forcing by a step function.

Let $f_{1,t}$ and $f_{2,t}$ be two solutions of the Loewner equation with a fixed pole and a unitary growth factor:
\begin{equation}
f_{1,t}(\omega) = \sqrt{ \omega^2 - 2\,t }
\end{equation}
and
\begin{equation}
f_{2,t}(\omega) = a_0 + \sqrt{ (\omega-a_0)^2 - 2\,t } \, .
\end{equation}
The first function $f_{1,t}$ is a solution of the Loewner equation with $a=0$, and is the identity mapping at $t=0$. Similarly, $f_{2,t}$ corresponds to a pole fixed at $a=a_0$ and is also the identity mapping at $t=0$.

The solution $f_d$ of the Loewner equation forced with the step function $a_d$ is $f_{1,t}$ before the discontinuity time $t_c$, and the composition of $f_{1,t}$ with $f_{2,t}$ later on:
\begin{equation}
f_{d,t} = \left\{
\begin{array}{ccc}
f_{1,t} & \mbox{if} & 0 \leq t < t_c \\ 
f_{1,t_c} \circ f_{2,t-t_c} & \mbox{if} & t \geq t_c
\end{array} 
\right. \, .
\label{eq:discontinuous_loewner}
\end{equation}
The discontinuity in pole's motion generates a new branch in the physical space. By virtue of the convolution, the new branch grows exactly along the flow line of the first mapping frozen at time $t_c$, $f_{1,t_c}$. As we reduce the amplitude of the step, the new branch nucleates closer and closer to the tip of the mother branch.

\subsection{Composition of slit mappings}

\begin{figure}
\centering
\includegraphics[width = 0.6\linewidth]{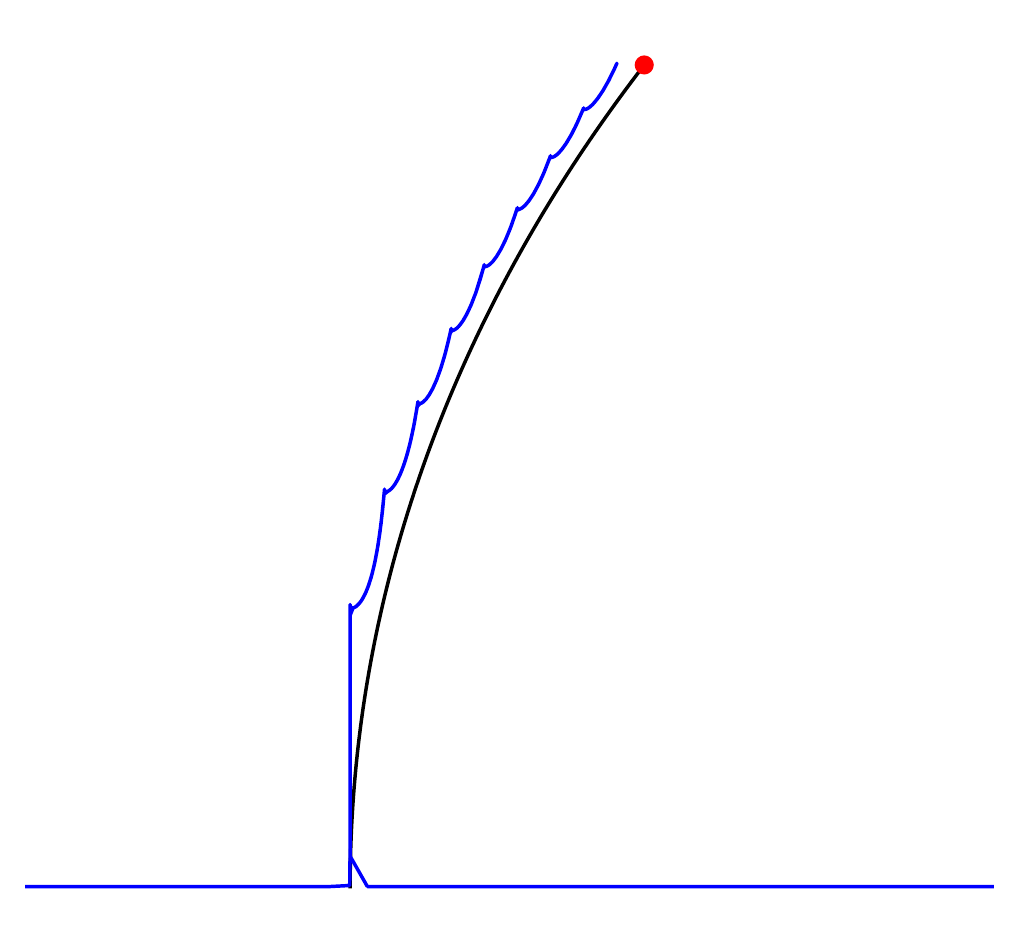}\\
\caption{Loewner growth of a finger with a linear forcing (black line, figure~\ref{fig:loewner_linear_forcing}), compared to its approximation by nesting a series of ten slit mappings (blue line).\label{fig:series_of_slit_maps}}
\end{figure}

In the above example, we have used the invariance of the Loewner equation with respect to composition (section~\ref{sec:discondinuous}). Based on this property, we can decompose the growth of a finger into a series of infinitesimal steps. Let $f_t$ be a solution of the Loewner equation at time $t$. We may write the solution at time $t+\delta t$ as
\begin{equation}
f_{t+\delta t} = f_t \circ \phi_{\delta t, a}
\label{compo}
\end{equation}
where $\phi_{\delta t, a}$ is the slit mapping:
\begin{equation}
\phi_{\delta t, a}(\omega) = a + \sqrt{ (\omega-a)^2 - 2\,\delta t } \, .
\label{slit}
\end{equation}
Indeed, the slit mapping satisfies the Loewner equation for a constant $a$, $f_t$ is a constant analytical function between $t$ and $t+\delta t$. %, and finally $\phi_{0, a}$ is the identity.
If we repeat this composition at each time step, the resulting map shows a series of off-tip growth which, collectively, divert the resulting finger from a straight line. As the time step $\delta t$ vanishes, the resulting map approaches the solution of the continuous Loewner equation (figure~\ref{fig:series_of_slit_maps}). The Loewner equation describing the growth of a finger is thus the continuous limit of a series of nested slit mappings. It is generally introduced as such in the literature  \cite{gruzberg2004loewner,gubiec2008fingered}.

So far, we have only considered the deterministic growth of a single finger. We next illustrate how random growth can create the branching fingers which constitute a proper network.

\subsection{Random forcing}

\begin{figure}
\centering
\includegraphics[width = \linewidth]{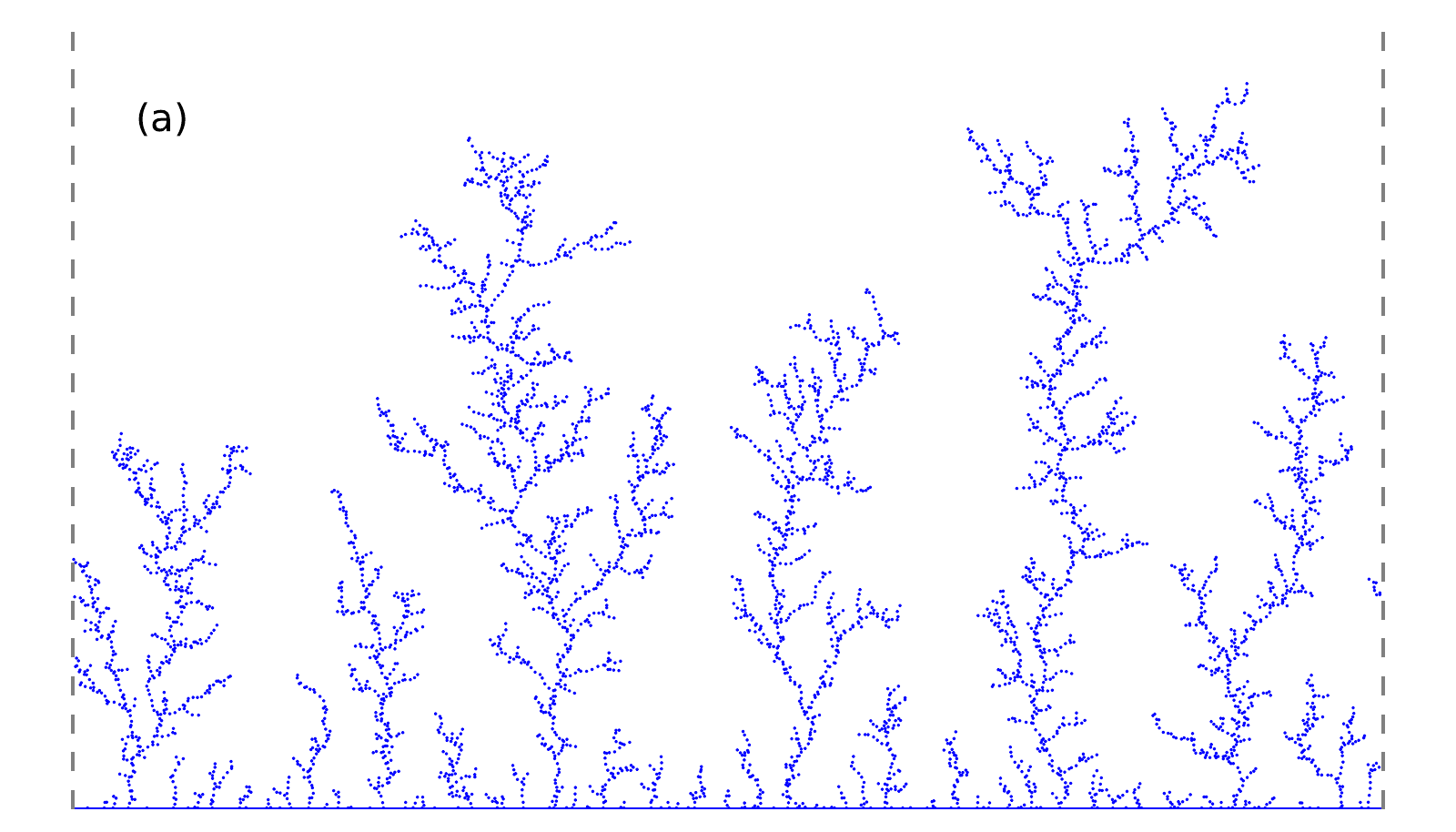}\\
\bigskip
\includegraphics[width = \linewidth]{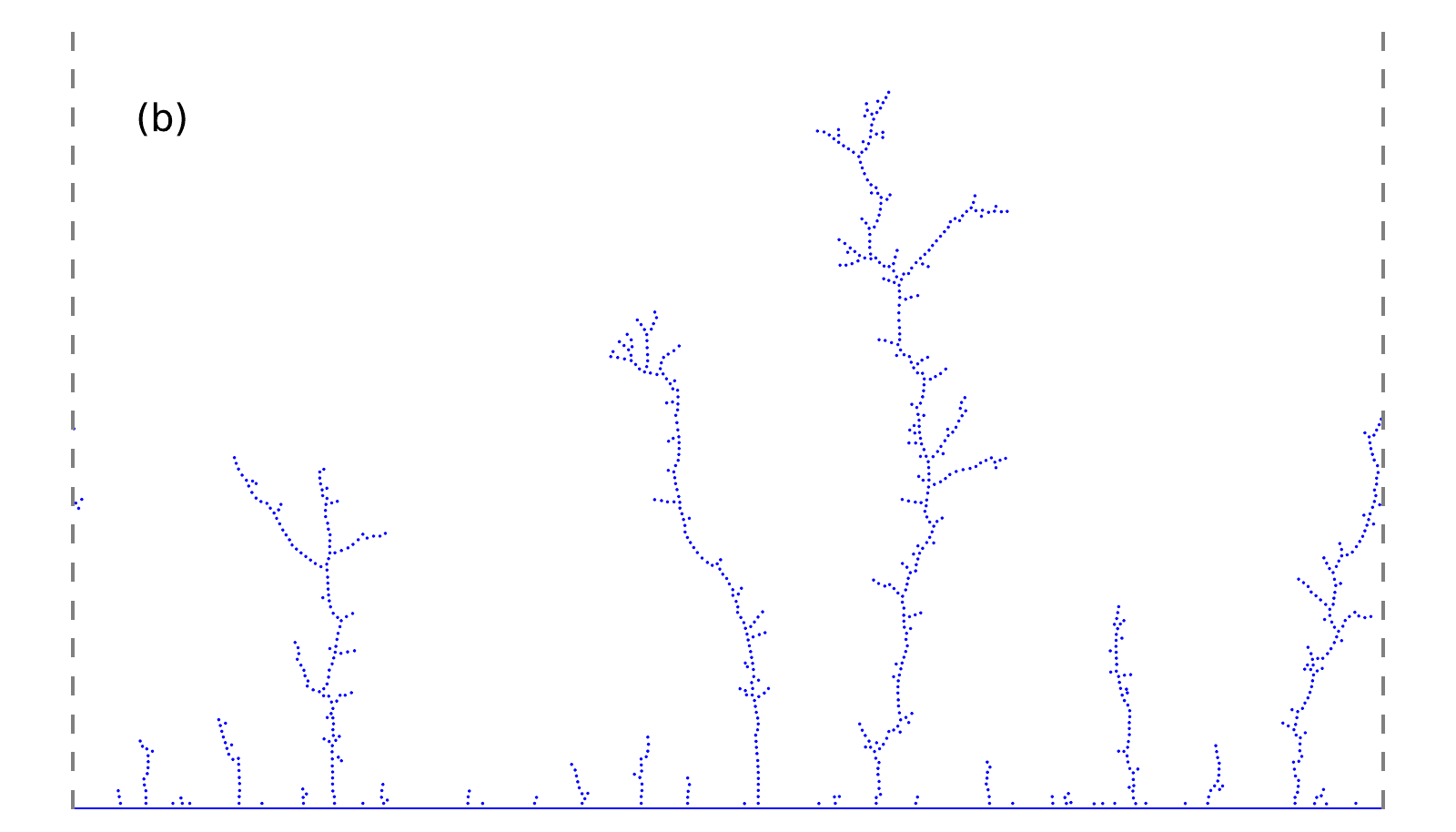}
\caption{Hastings-Levitov growth. (a) $\eta=1$ (Laplacian growth), (b) $\eta=1.8$. Dashed lines indicate periodic boundaries.\label{fig:HL}}
\end{figure}

Laplacian networks often result from intermittent processes involving some degree of randomness. For instance, the nucleation of a new crack in a brittle material, that of a dendrite on a growing crystal, or the exact direction in which a spring erodes the landscape, are all controlled by specific conditions at the microscopic scale. At the scale of the network however, Laplacian growth sometimes contracts complicated microscopic dynamics to more deterministic rules, while still allowing for the creation of new branches.

The emergence of the network's macroscopic properties from random growth has attracted much attention. In the Schramm-Loewner evolution, for instance, the pole $a$ performs a random walk along the real axis, and the corresponding finger draws a random curve which never intersects itself \cite{Lawler:2004,Kager:2004,Bauer:2006}. The growth mechanism introduced by Hastings and Levitov  is more closely related to our subject: a series of randomly-located infinitesimal growth events can generate a network of well-defined fingers \cite{Hastings:1998,Hastings2001,Hastings2001b}. The path that connects random and intermittent growth to the continuous advance of smooth fingers remains to be fully explored. Here, to set the stage for our subsequent discussion of interactions between the branches of a growing network, we simply illustrate how a macroscopic network can emerge from random microscopic growth in the Hasting-Levitov model.

We consider a model of diffusion-limited aggregation (DLA), where the position of the pole $a$ is randomly sampled from some distribution \cite{Hastings:1998}. Specifically, if the growth is driven by the harmonic field $p$, the probability $\mathrm{d}P$ that a walker joins the cluster within the interval $\mathrm{d}s$ of the boundary is $\mathrm{d}P = | \nabla p | \, \mathrm{d}s$. Since the harmonic measure is uniform in the mathematical plane, one can simply sample the position of the pole from a uniform distribution each time a new walker joins the cluster. The growth corresponds to the composition of the mapping $g$ with an elementary map, as in equation \eqref{compo}. The iteration of this procedure generates an aggregate in the physical plane, whose fractal dimension is that of a DLA cluster (figure~\ref{fig:HL}) \cite{Hastings:1998}.

A relatively straightforward generalization of this algorithm is to require the growth factor to be proportional to some power $\eta$ of the harmonic measure~\cite{Hastings2001b,Vicsek}. In the mathematical plane, the average number of walkers joining the cluster at a specific location is then proportional to $|f'|^{1-\eta}$. As $\eta$ increases, the probability of generating a new tip concentrates near existing tips, and the fractal dimension of the aggregate decreases.
More specifically, at each time step, the slit mapping \eqref{slit} introduces a new singularity in $g$, since $f' \sim \omega - a $ near the pole $a$. The probability measure thus becomes non-integrable when $\eta$ gets larger than 2 (this value actually depends on the infinitesimal map \cite{Hastings2001,Hastings2001b,Mathiesen2008}). The walkers then accumulate on existing tips and the aggregate becomes non-fractal. Instead, the DLA cluster grows continuous fingers, much like the continuous Loewner equation. Just before this transition, however, new fingers still nucleate from time to time to form a ramified network (figure~\ref{fig:HL}).

At this point, it is tempting to describe the growth of such a network as a continuous process, disrupted only by branching events. To do so, we need to establish macroscopic rules for the velocity and direction of the growth, and the nucleation of new fingers. Such rules can either be fully deterministic, or inherit some randomness from the microscopic growth process. Deriving them rigorously from the latter is certainly a formidable task, much like deriving them from basic principles in a physical system.

A more modest approach is to investigate the geometrical consequences of some simple rules, and identify their signature in the network they induce. In the next sections, we return to the continuous Loewner equation, and show how the rule controlling the growth direction relates to the geometry of interacting fingers.

\subsection{Multiple fingers}

\begin{figure}
\centering
\includegraphics[width = 0.6\linewidth]{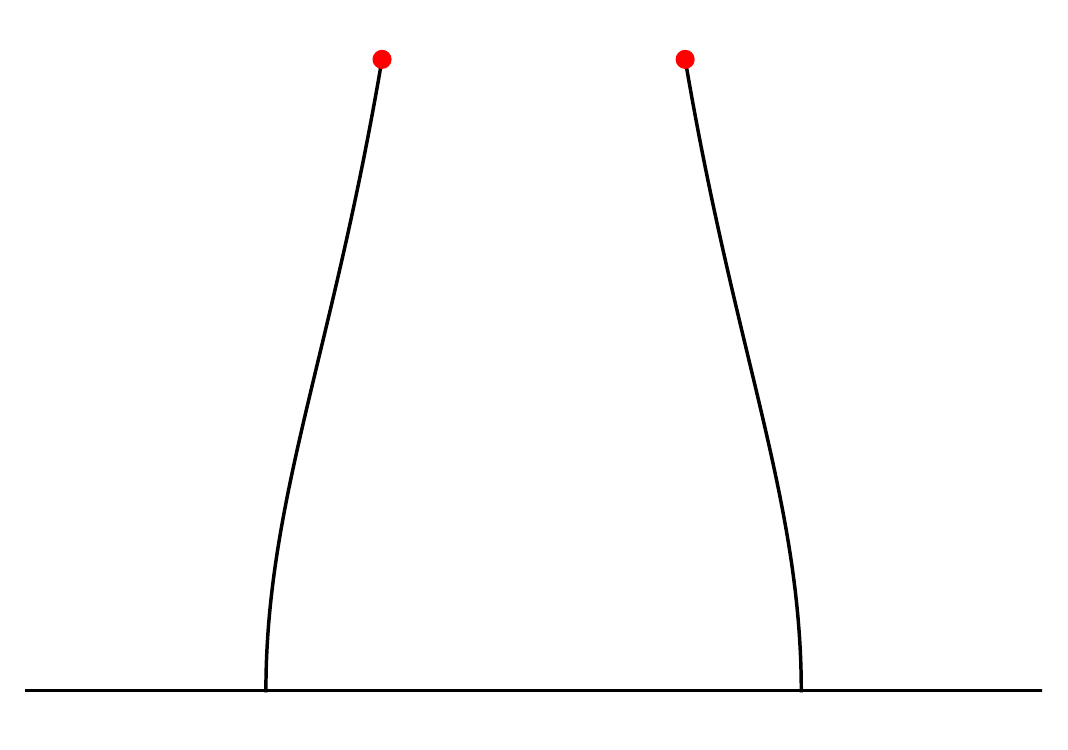}
\caption{Two fingers growing in the same diffusion field, with static poles and constant growth factors.\label{fig:two_needles_not_geometric}}
\end{figure}

An interesting property of Laplacian networks is the interaction among its fingers via the diffusion field \cite{derrida1992needle}. This collective behavior is naturally encoded in the Loewner equation. To illustrate this property, let us consider two fingers growing in the same diffusion field. Following the reasoning of section~\ref{sec:growth}, the associated growth function shares its two poles $a_1$ and $a_2$ with the inverse map :
\begin{equation}
\dfrac{1}{\varphi} = \dfrac{G_1}{\omega - a_1} + \dfrac{G_2}{\omega - a_2}
\label{eq:two_needles_growth_function}
\end{equation}
where the growth factors $G_n$ and the poles $a_n$ are real numbers. This expression for $\varphi$ satisfies boundary condition \eqref{eq:far_field_BC} at infinity and its imaginary part vanishes on the real axis.

The simplest possible rule for the motion of the poles is to keep them fixed as the fingers grow (without loss of generality, $a_1 = 1$ and $a_2=-1$). Interpreting the position of a pole as the value of the stream function at the corresponding tip, we find an immediate consequence of this rule: the value of the stream function at the first tip departs from its value at the other one by $2$. This property has a rather peculiar interpretation: it means that the growing fingers maintain the total flux passing between their tips. We nonetheless use it to illustrate the interaction between fingers.

For simplicity, we also assume constant growth factors ($G_1=G_2=1$). Using again the invariance of the Loewner equation with respect to composition, we find that the solution which initially is the identity reads
\begin{equation}
f_t(\omega) = h^{-1} \left( h(\omega) - t \right)
\end{equation}
where the mapping $h$ is
\begin{equation}
h(\omega)=\dfrac{\omega^2}{4}- \dfrac{\ln(\omega)}{2}
\end{equation}
and $h^{-1}$ is its inverse \cite{kager2004exact}.

The two symmetric fingers generated by this simple example bend toward each other as they grow (figure~\ref{fig:two_needles_not_geometric}), a behavior we interpret in the following way. As they grow off of the real axis, the fingers are able, collectively, to catch more and more flux. To compensate for this increase, they grow closer to each other, in a way that maintains the flux between their tips.

In contrast with the straight growth of a single finger forced with a static pole (section~\ref{sec:static_pole}), here the growth of each finger influences the shape of the other. This interaction, mediated by the diffusion field, vanishes when the fingers are far from each other. Initially, each finger therefore grows orthogonally to the real axis, until the other's presence becomes perceptible. In section~\ref{sec:max_flux}, we will turn this interaction into competition for the diffusive field.

The growth of an arbitrary number of fingers is straightforwardly generalized from equation (\ref{eq:two_needles_growth_function}): it is the sum of as many poles as there are growing tips \cite{hastings2001growth,carleson2002laplacian}. The corresponding growth function reads
\begin{equation}
\dfrac{1}{\varphi} = \sum_{n=1}^N \dfrac{G_n}{\omega - a_n}
\label{eq:N_needles_phi}
\end{equation}
where $N$ is the total number of fingers. The above formula encapsulates the three rules that define the growth of a network in a diffusion field. The velocity of each tip is controlled by the associated growth factor $G_n$, its direction is governed by the motion of the pole $a_n$ along the real axis, and the nucleation of a new tip translates into the appearance of a new addend in equation \eqref{eq:N_needles_phi}.

These three rules mark the shape of the network they generate. Hereafter, we focus on different aspects of the second growth rule, which controls the direction of a finger.

\section{Growth rules}\label{sec:growth_rules}

The previous section has illustrated the relation of a pole's motion to the growth of fingers. Here we outline the importance of a growth rule that drives a finger along the flow lines of a Laplacian field. Following Carleson and Makarov \cite{carleson2002laplacian}, we refer to these dynamics as ``geodesic growth''.
We begin with a recipe for geodesic growth \cite{selander,carleson2002laplacian, gubiec2008fingered}. We then propose a new perspective on the recently introduced analogy between geodesic growth and the principle of local symmetry of fracture mechanics (PLS) \cite{barenblatt1961brittle,gol1974brittle, yossi}. Finally, we show the relation between the PLS and some optimization principles in the context of Laplacian networks.

\subsection{Geodesic growth}\label{sec:geo_growth}

\begin{figure}
\centering
\includegraphics[width = 0.49\linewidth]{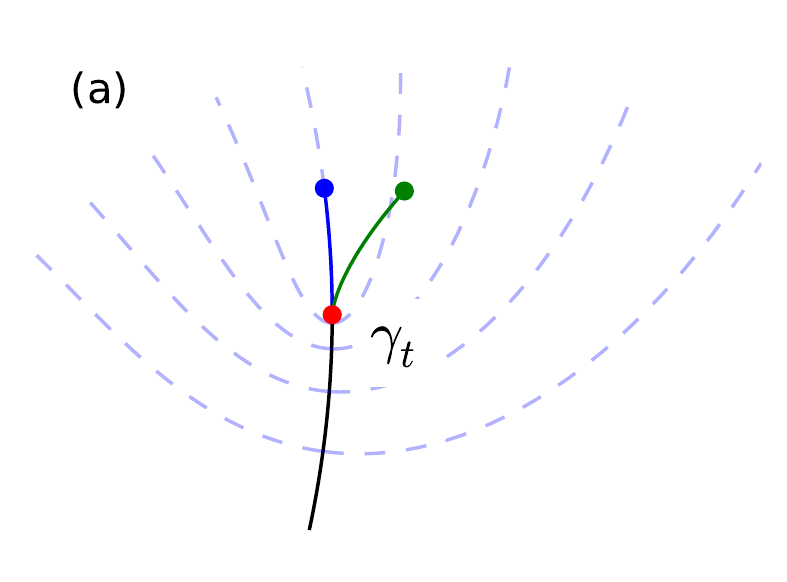}
\includegraphics[width = 0.49\linewidth]{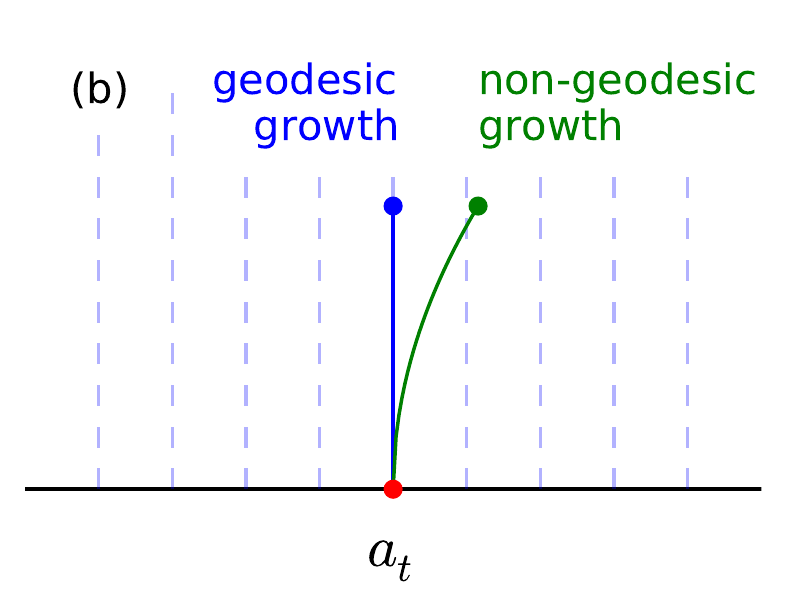}
\caption{Definition of geodesic growth in the physical plane (a) and in the mathematical plane (b). This sketch represents growth up to order $\delta t$ only (equation~(\ref{eq:parametrize_tip_motion})); at higher order, even a geodesic finger can bend away from the flow line corresponding to time $t$. \label{fig:geodesic_growth}}
\end{figure}

The Loewner equation naturally grows smooth fingers when the forcing (i.e. the growth factor and the pole motion) is smooth. Indeed, following equation (\ref{eq:gamma_dot_G}), the velocity of the tip in the physical plane is tangential to the finger's tip. Nonetheless, by moving in the mathematical plane, the corresponding pole controls the curvature of the tip in the physical plane, and thus the direction in which it is heading. The motion of the pole, like the intensity of the growth factor, depends on the microscopic details of the system we consider; both rules must be specified based on physical considerations.

To our knowledge, only two deterministic rules have been invoked in the context of Laplacian growth: straight and geodesic growth. Straight needles can represent growing crystals, whose shape is constrained by the regular arrangement of atoms or molecules \cite{derrida1992needle}, or viscous fingers invading a regular network of channels  \cite{pecelerowicz2014effective}. Although geodesic growth is common in the literature, its physical origin is seldom investigated \cite{selander,carleson2002laplacian,gubiec2008fingered,yossi}.

A geodesic finger grows along the flow line intersecting its tip (figure~\ref{fig:geodesic_growth}). For instance, a straight finger growing in the upper half plane with a fixed pole is geodesic (equation (\ref{eq:strait_needle})). However, geodesic growth does not necessarily produce straight fingers: after the discontinuity occurring at time $t_c$, the Loewner growth corresponding to equation (\ref{eq:discontinuous_loewner}) is geodesic, and nevertheless produces a curved finger (figure~\ref{fig:discontinuous_growth}).

Similarly, the geodesic growth of a tip does not require the pole to be fixed in the mathematical plane. Rather, it means that the growing finger remains on the streamline that went trough its tip an instant $\delta t$ before (figure~\ref{fig:geodesic_growth}). In the mathematical plane, this condition means that the trajectory of the tip's image is a vertical straight line \cite{selander}. We now formalize this definition by considering the image of the growing tip at time $t + \delta t$ in the mathematical plane corresponding to time $t$, that is  $g_t(\gamma_{t+\delta t})$.

As suggested by equation (\ref{eq:strait_needle}), we first parametrize the motion of the tip in the mathematical plane with $\sqrt{\delta t}$ \cite{gubiec2008fingered}:
\begin{equation}
g_t ( \gamma_{t+\delta t} )= a_t + \alpha \sqrt{\delta t} + \beta \delta t + {\cal O} \left( \delta t^{3/2} \right)
\label{eq:parametrize_tip_motion}
\end{equation}
where $\alpha$ and $\beta$ are complex numbers. The parameter $\alpha$ encodes the growth direction of the finger. If it is a pure imaginary number, the finger grows perpendicularly to the real axis in the mathematical plane, and therefore grows along the flow line in the physical plane. The parameter $\beta$ controls its curvature. Next we derive expressions for $\alpha$ and $\beta$ from the Loewner equation.

The mapping $g_t$ is singular at $\gamma_{t}$, but we may expand it up to first order in time near $\gamma_{t+\delta t}$:
\begin{equation}
g_t = g_{t+\delta t} - \delta t \, \dot{g}_{t}  + {\cal O} \left( \delta t^2 \right) \, .
\label{eq:g_time_expansion}
\end{equation}
Using the Loewner equation (\ref{eq:loewner_inverse}), which relates the time derivative of the mapping to the growth function $\varphi$, and applying the above expansion to $\gamma_{t+\delta t}$, we find
\begin{equation}
g_t( \gamma_{t+\delta t} ) = a_{t+\delta t} - \dfrac{\delta t} { \varphi \circ g_{t}( \gamma_{t+\delta t} )}  + {\cal O} \left( \delta t^2 \right)
\label{eq:g_time_expansion_loewner}
\end{equation}
where $\varphi$ is evaluated at time $t$.

Finally, we substitute $g_t( \gamma_{t+\delta t} )$ for its parametrization (\ref{eq:parametrize_tip_motion}) in equation (\ref{eq:g_time_expansion_loewner}), and expand the result up to order $\delta t$. Matching the orders of the resulting expansion, we find at first order
\begin{equation}
\alpha = \dfrac{i}{\sqrt{\varphi'(a)}} \, .
\end{equation}
Thus, $\alpha$ is a pure imaginary number, and the finger grows along the flow line at first order, regardless of the pole's motion. This is consistent with equation~(\ref{eq:gamma_dot_G}). At next order, we find 
\begin{equation}
2\beta = \dot{a} + \dfrac{\varphi''(a)}{2 \varphi'(a)^2} \, .
\label{eq:beta_loewner}
\end{equation}
According to this equation, the motion of the pole $a$ controls the value of the parameter $\beta$, and therefore the bending of the finger away from the flow line.

The Loewner equation produces fingers that grow, at first order, along the flow line intersecting their tip. Geodesic growth further require them to remain along the flow line at next order. Mathematically, this means that the real part of $\beta$ vanishes (figure~\ref{fig:geodesic_growth}). Accordingly, equation~(\ref{eq:beta_loewner}) yields a formal definition of geodesic growth for the Loewner equation \cite{yossi}:
\begin{equation}
\dot{a} = - \frac{\varphi''(a)}{2\,\varphi'(a)^2} \, .
\label{eq:geodesic_rule}
\end{equation}
This differential equation controls the motion of the pole $a$ for a geodesic finger. When the coefficient $\varphi''(a)$ vanishes, the geodesic growth of a finger implies that the corresponding pole is fixed in the mathematical plane (equations (\ref{eq:strait_needle}) and (\ref{eq:discontinuous_loewner}), for instance), but this is not true in general.

\begin{figure}
\centering
\includegraphics[width = 0.49\linewidth]{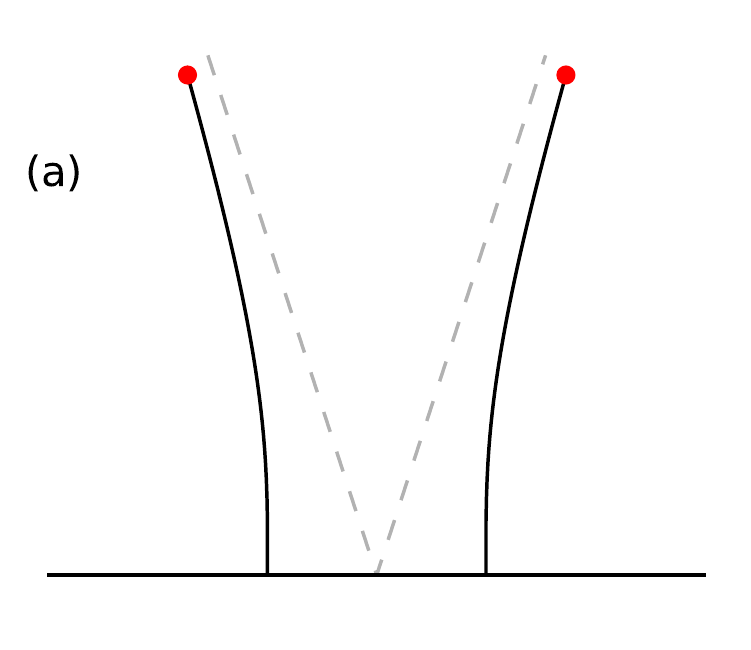}
\includegraphics[width = 0.49\linewidth]{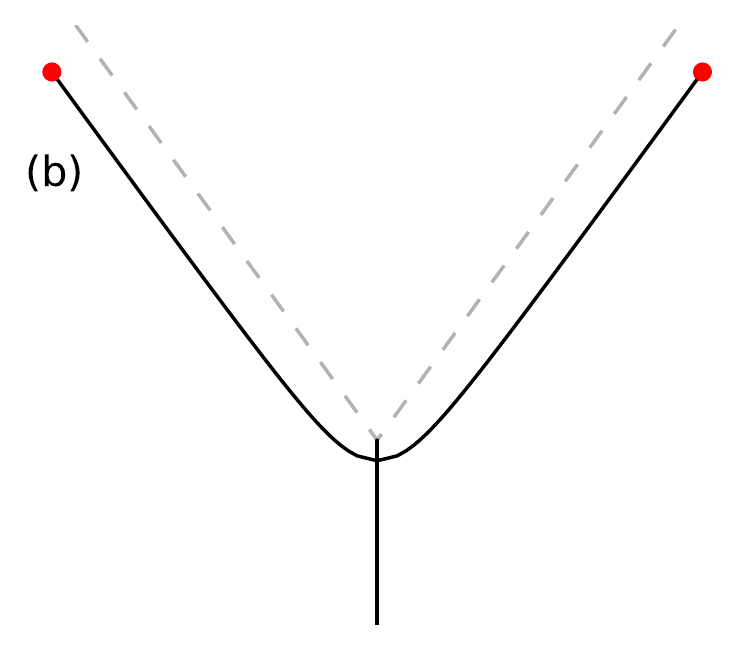}
\caption{Geodesic growth of two symmetric fingers off of the real axis, at time $t=3$ with growth factor $G=1$ (a). The poles are located initially at -1/2 and 1/2, and the aspect ratio is preserved. Dashed lines are asymptotes, with an opening angle of $\pi/5$. Same figure mapped with the square function (b). The opening angle of the asymptotes is $2\pi/5$.\label{fig:geodesic_growth_two_tips}}
\end{figure}

When geodesic fingers grow in the same diffusion field, their growth moves the poles in the mathematical plane, as each one perturbs the trajectory of the others. For illustration, let us consider two geodesic fingers, symmetric with respect to the imaginary axis. The two corresponding poles are located at
\begin{equation}
a_- = - a(t)
\quad \mbox{and} \quad
a_+ = a(t)
\, .
\end{equation}
After equation (\ref{eq:N_needles_phi}), and assuming that the growth factor is $G=1$ for both fingers, the growth function reads
\begin{equation}
\varphi = \dfrac{\omega^2 - a^2}{2\,\omega} \,.
\end{equation}
Applying the geodesic rule (\ref{eq:geodesic_rule}) to each pole yields
\begin{equation}
\dot{a}= \dfrac{1}{2\, a} \, .
\end{equation}
Therefore, the two poles repel each other and move apart along the real axis of the mathematical plane:
\begin{equation}
a = \sqrt{t+a_0^2}
\end{equation}
where $a_0$ sets the initial position of the poles.  We then integrate the Loewner equation (\ref{eq:loewner_inverse}) to find the mapping $g$ \cite{gubiec2008fingered}:
\begin{equation}
5 g^{1/2} a^2 - g^{5/2} = 5 g_0^{1/2} a_0^2 - g_0^{5/2}
\label{eq:integrated_loewner}
\end{equation}
where the mapping $g_0$ defines the initial geometry of the growing fingers. At a tip ($z=\gamma_{\pm}$), and when $g_0$ is the identity, the above equation implicitly determines the trajectories of the tips:
\begin{equation}
\gamma_{\pm} ( \gamma_{\pm}^2 - 5\,a_0^2 )^2 = \pm 16 ( a_0^2 + t )^{5/2} \, .
\label{eq:geodesic_gamma}
\end{equation}
The two fingers, as they compete for the available field, bend out to avoid each other (figure~\ref{fig:geodesic_growth_two_tips}). At long times, they asymptotically tend towards two straight lines with an opening angle of $\pi/5$. This behavior contrasts with the growth of symmetric fingers with fixed poles (figure~\ref{fig:two_needles_not_geometric}): here, the geodesic growth directs the fingers towards free space, where more diffusive field is available.

Similarly, the behavior of a collection of $N$ geodesic fingers translates into a system of $N$ ordinary differential equations controlling the motion of their associated poles. Indeed, applying the geodesic rule (\ref{eq:geodesic_rule}) to each pole, together with the general expression of the growth function (\ref{eq:N_needles_phi}), we find \cite{gubiec2008fingered}
\begin{equation}
\dot{a}_m = \sum_{n=1, \, n\neq m}^N \dfrac{G_n}{a_m - a_n}
\label{eq:poles}
\end{equation}
for any pole labeled by $m$. Thus, the motion of the poles is, in general, coupled to the distance between them. This coupling, of course, is affected by the growth factors, which can also depend on the location of the poles.

The above reasoning applies to fingers growing into the upper half plane, but similar equations have been established in radial, annular or cylindrical geometries \cite{Bauer:2006,gubiec2008fingered,Zhan2004}.

\subsection{Principle of local symmetry} \label{sec:PLS}

When a quasi-static fracture propagates in an elastic body, it is sometimes postulated to follow the path which maintains a symmetric deformation around its tip \cite{barenblatt1961brittle}. This growth rule is known as the principle of local symmetry (PLS). It has been recently invoked to represent the growth Laplacian fingers, based on its equivalence with geodesic growth \cite{yossi}.

To understand this equivalence, we need to formalize the concept of local symmetry. We first expand the map $f$ around a pole, and interpret its coefficients :
\begin{equation}
f(\omega) = \gamma + c_2 \, (\omega -a)^2 + c_3 \, (\omega -a)^3 + {\cal O}(\omega -a)^4 
\label{eq:map_expansion}
\end{equation}
The first coefficient $\gamma$ is the position of the tip corresponding to the pole $a$. The orientation of this tip is the argument of the second-order coefficient $c_2$.

The magnitude of $c_2$ represents the intensity of the diffusion flux in the tip's neighborhood. For instance, if the finger epitomizes a branch generated by diffusion-limited aggregation (DLA), the number of random walkers joining the tip per unit time is proportional to $\left| c_2 \right|^{-1/2}$, and so is the velocity at which the tip grows \cite{derrida1992needle,gubiec2008fingered}.

\begin{figure}
\centering
\includegraphics[width = 0.49\linewidth]{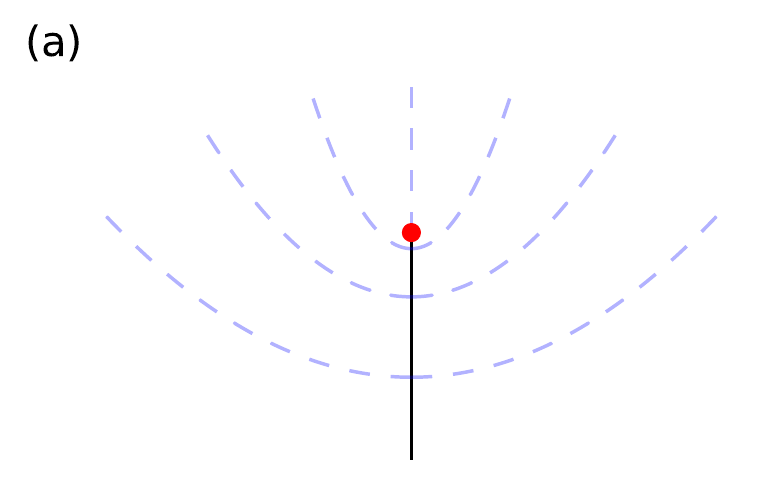}
\includegraphics[width = 0.49\linewidth]{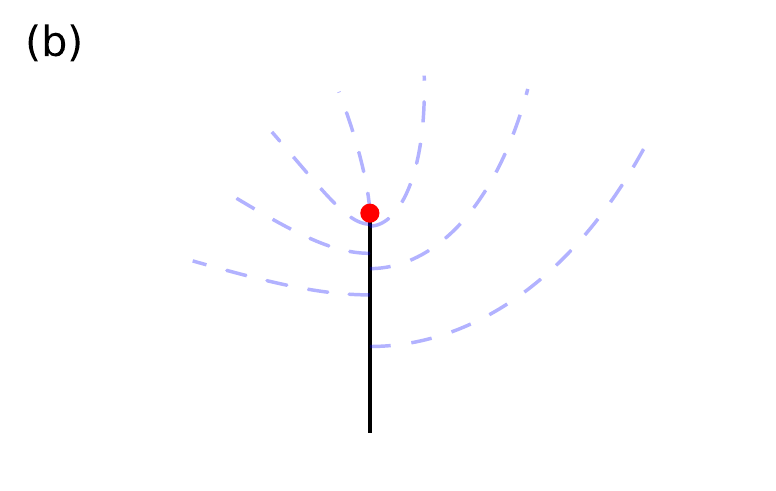}
\caption{Flow lines around a tip for $c_2=-i$. The flow is locally symmetric when the third order coefficient $c_3$ vanishes (a). A finite value of $c_3$ breaks this symmetry, $c_3=-0.3\, i$ (b).\label{fig:local_symmetry}}
\end{figure}

Truncating the expansion of the map $f$ at second order, we find a symmetric diffusion field (figure~\ref{fig:local_symmetry}). The third-order coefficient $c_3$ breaks this symmetry \cite{petroff2013bifurcation}. When it is finite, the flow lines bend out of alignment with the tip. In the case of DLA, it means that more walkers come from one side of the tip than from the other. Based on this observation, we define the PLS in the context of Laplacian networks as $c_3=0$. As it grows according to this definition, a finger maintains, up to third order, a symmetric diffusion field around its head. In general, this property does not extend to higher orders, which makes the PLS only local. We now establish its equivalence with geodesic growth.

Inserting expansion (\ref{eq:map_expansion}) into the Loewner equation (\ref{eq:fpfd}), and expanding again the result near the pole yields, at first order,
\begin{equation}
2 \, c_2 = - \varphi'(a) \, \dot{\gamma} \, .
\label{eq:first_order_expansion}
\end{equation}
This expression simply summarizes the continuous growth of a finger: the tip velocity $\dot{\gamma}$ is aligned with the tip itself. The next order of the same expansion reads
\begin{equation}
3\,c_3 = - \dot{\gamma} \left( \dfrac{\varphi''(a)}{2} + \dot{a} \, \varphi'(a)^2 \right) \, .
\end{equation}
Therefore, whenever the tip grows ($\dot{\gamma} \ne 0$), the PLS ($c_3=0$) is equivalent to geodesic growth (equation (\ref{eq:geodesic_rule})).

While mathematically equivalent to geodesic growth, the PLS has a distinct physical interpretation: is is a geometrical rule, rather than a dynamical one. Indeed, the PLS states that the shape of a network of geodesic fingers must be such that the diffusive field is locally symmetric around each tip. For instance, we can recognize the couple of fingers represented in figure \ref{fig:two_needles_not_geometric} as non-geodesic by solving for the Laplacian field around them, and observe that the third coefficient of its expansion around each tip does not vanish.

\begin{figure}
\centering
\includegraphics[width = 0.49\linewidth]{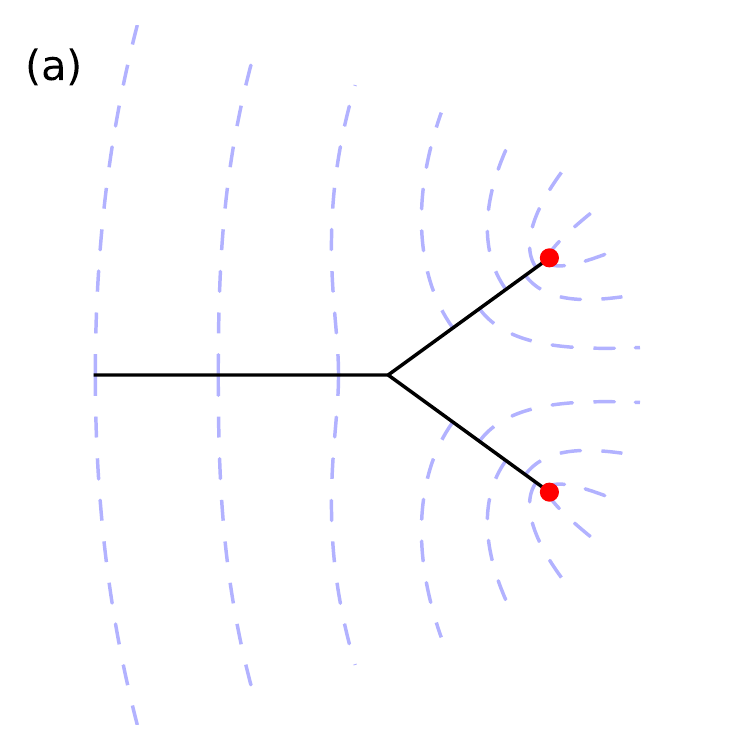}
\includegraphics[width = 0.49\linewidth]{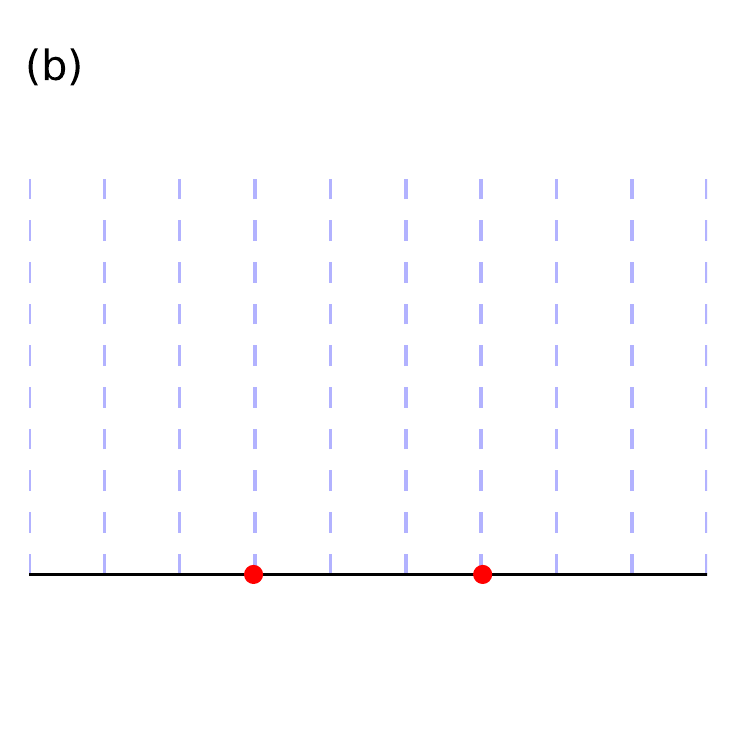}
\caption{Bifurcating finger in the physical plane (a) and after a mapping (\ref{eq:bifurcation_map}) to the mathematical plane (b). Dashed blue lines: flow lines from mapping (\ref{eq:bifurcation_map}). The opening angle between the daughter branches is $ 2 \pi / 5$ in this example. \label{fig:bifurcation_map}}
\end{figure}

Bifurcations make for a more instructive example of the correspondence between growth dynamics and geometry \cite{carleson2002laplacian,devauchelle2012ramification,petroff2013bifurcation}. Indeed, we can invoke the PLS to calculate the angle of a symmetric geodesic bifurcation. Let us consider two straight branches of length unity, emanating from their mother branch at an angle $ 2 \pi c$ (figure~\ref{fig:bifurcation_map}). The conformal map $f_{c}$ brings the region outside the fingers to the mathematical plane:
\begin{equation}\label{eq:bifurcation_map}
%f_{c}( \omega ) =  \omega^{2c} \left(\omega^2-\dfrac{a^2}{c} \right)^{1 - c}
f_{c}( \omega ) = \omega^{2c} \left( \left( \omega - \dfrac{a}{\sqrt{c}} \right) \left( \omega + \dfrac{a}{\sqrt{c}}\right) \right)^{1 - c}
\end{equation}
The two poles corresponding to the two tips are located at $\pm a$. The principle of local symmetry simply reads $f_c'''(a)=0$, which yields
\begin{equation}
\frac{4(1 - 5c)c^c}{a (1-c)^{c+1}} = 0
\end{equation}
and consequently, $c=c_{\mathrm{PLS}}=1/5$. Accordingly, the bifurcation angle corresponding to local symmetry is $2\pi/5=72^{\circ}$. Previous contributors derived this finding from different considerations \cite{Hastings2001,carleson2002laplacian,devauchelle2012ramification,petroff2013bifurcation}.

This special angle is related to the geodesic growth of two fingers off of the real axis (section~\ref{sec:geo_growth}). The invariance of the Loewner equation with respect to composition allows us to map this system onto an approximate bifurcation with daughter branches starting slightly behind the tip of the mother branch (figure~\ref{fig:geodesic_growth_two_tips}). The daughter branches bend towards the $2\pi/5$ bifurcation, suggesting that this shape is an attractor for geodesic growth. When the bifurcation is symmetric and the growth factor positive, as in the present example, the $2\pi/5$ bifurcation is indeed a stable fixed point \cite{selander}. As such, this special angle can be interpreted as the geometrical signature of geodesic growth in a diffusive field.

\subsection{Maximization of the flux entering the tip} \label{sec:max_flux}

Fracture mechanics inspired the introduction of the PLS in the context of Laplacian networks. Can this analogy shed more light on the growth of Laplacian fingers? In linear-elasticity fracture mechanics, one often assumes that a crack follows a trajectory that maximizes the energy release rate \cite{barenblatt1961brittle,gol1974,freund1998,cohen2010}, although the non-linearity of the the stress field can induce cracks to depart from this rule \cite{Amestoy1992,adda2005brittle}. In most systems, this optimal growth is equivalent to the PLS \cite{cotterell1980}. We now ask whether a similar optimization principle could direct the growth of Laplacian fingers.

When a geodesic finger grows along a flow line, it bends towards the strongest source of diffusive flux. Our intuition suggests that, in doing so, it would maximize the flux entering its tip. To formalize this idea, we first need to evaluate the change of flux associated with an infinitesimal increase $\vert\mathrm{d} \gamma\vert$ of the finger's length. Then, we will identify the growth direction that maximizes the increase of the flux into the tip.

In the tip's neighborhood, this flux is proportional to $\left| c_2 \right|^{-1/2}$ (section~\ref{sec:PLS}). Its rate of change as the finger grows therefore reads
\begin{equation}
\dfrac{\mathrm{d} \left( \vert c_2 \vert^{-1/2} \right) }{\vert \mathrm{d} \gamma \vert} =
 - \dfrac{1}{2 \left| c_2 \right|^{3/2}} \dfrac{\mathrm{d} \vert c_2 \vert}{\vert \mathrm{d} \gamma \vert} 
\label{eq:maxi_1}
\end{equation}
Being normalized with the finger's length increase $\vert\mathrm{d} \gamma\vert$, this quantity does not depend on the tip's velocity. We now determine how it depends on the growth direction.

The mapping $f$ evolves according to the Loewner equation (\ref{eq:fpfd}), and so does its expansion near the pole, including the coefficient $c_2$.
%\begin{equation}
%\dfrac{\mathrm{d}\vert c_2 \vert}{\mathrm{d}t} = \vert c_2 \vert \, \re \left(\dfrac{\dot{c_2}}{c_2}\right) \, ,
%\end{equation}
Using equation (\ref{eq:first_order_expansion}), which derives from the Loewner equation, we find
\begin{equation}
\dfrac{\mathrm{d}\vert c_2 \vert}{\vert \mathrm{d} \gamma \vert} = \dfrac{ \left| \varphi'(a) \right| }{4} \, \re \left(\dfrac{\dot{c}_2}{c_2}\right) \, .
\label{eq:maxi_2}
\end{equation}

Next, we express $\dot{c}_2$ in terms of the pole velocity $\dot{a}$. To do so, we need to expand the growth function $\varphi$ and the mapping $f$ around the pole up to third and fourth order respectively. Substituting these expansions into the Loewner equation, and using equation (\ref{eq:first_order_expansion}) again, we finally obtain
\begin{equation}
\re \left(\dfrac{\dot{c}_2}{c_2}\right)  = 
%- \dfrac{4 c_4}{ c_2 \varphi'} + \dfrac{\varphi'''}{3 \varphi'^2} + 2 \dot{a} \left( \dfrac{\varphi''}{\varphi'} + \dot{a}\,\varphi' \right)
2 \varphi' \dot{a}^2 + 2 \dfrac{\varphi''}{\varphi'} \dot{a}  + \dfrac{\varphi'''}{3 \varphi'^2} - \re \left( \dfrac{4 c_4}{ c_2 \varphi'} \right)
\label{eq:maxi_3}
\end{equation}
where all derivatives of $\varphi$ are evaluated in $a$, where they are real.

At a given time during the growth of a finger, all derivatives of $f$ and $\varphi$ are set, but we may still chose the velocity of the pole $\dot{a}$. Combining equations (\ref{eq:maxi_1}), (\ref{eq:maxi_2}) and (\ref{eq:maxi_3}), we note that the rate of change of the flux into the tip is a second-order polynomial of the free parameter $\dot{a}$, which we can optimize straightforwardly. We then find that $\dot{a}$ follows the geodesic growth rule (\ref{eq:geodesic_rule}).

The sign of $\varphi'(a)$ decides whether the geodesic growth maximizes the flux into the tip, or minimizes it. After equation (\ref{eq:phi_near_pole}), $\varphi'(a)$ is the inverse growth factor of the tip corresponding to the pole $a$; it is therefore positive for an advancing tip. As a consequence, a geodesic finger grows in the direction that maximizes the diffusive flux into its tip. It is therefore equivalent to say that a finger grows geodesically, that it satisfies the principle of local symmetry, or that it maximizes the flux into its tip. These rules are just three facets of the same dynamics, and we use them interchangeably hereafter.

To state it informally, a geodesic finger makes the most of its growth. As it tries to maximize the flux into its tip, however, its growth sometimes causes the flux to decline. Indeed, substituting the geodesic growth condition (\ref{eq:geodesic_rule}) into equations (\ref{eq:maxi_1}), (\ref{eq:maxi_2}) and (\ref{eq:maxi_3}), we find that the variation of the flux reads
\begin{equation}
\dfrac{\mathrm{d} \left( \vert c_2 \vert^{-1/2} \right) }{\vert \mathrm{d} \gamma \vert} =
\dfrac{1}{2 \left| c_2 \right|^{3/2}}
\left(
\dfrac{\varphi''^2}{8\varphi'^2}
-\dfrac{\varphi'''}{12\varphi'}
+ \re \left( \dfrac{c_4}{c_2} \right)
\right)
\label{eq:maxi_4}
\end{equation}
for a geodesic finger. This quantity can be negative, indicating that the flux can decrease as the tip advances. Although optimal for a fixed increase of the finger's length, the geodesic growth can bring the finger away from a more advantageous configuration. In the next section, we illustrate this apparent paradox by considering the growth of a symmetric bifurcation.

\section{Optimal branching angles}\label{sec:opti_branch}

In section~\ref{sec:max_flux}, we formulated an optimal rule for the growth of geodesic fingers. This formulation prompts us to investigate the optimality of the networks such fingers generate. Drainage networks are often thought to be optimal with respect to some global quantity \cite{howard1990theoretical,Rinaldo1992,maritan1996universality}. Is there a global quantity that a network of geodesic fingers optimizes as it grows? We are not able to answer this question in general\footnote{When all growth factors are constant, there exists a Lyapunov function that geodesic growth minimizes \cite{selander}.}.

In this section, we investigate a modest network, consisting of a symmetric bifurcation, from the standpoint of flux optimization. First we adapt the method of Section~\ref{sec:geo_growth} to calculate the trajectories of two symmetric fingers growing off of an initially straight bifurcation. We then use this result to represent the fingers' trajectory in terms of flux and orientation, and compare these optimal trajectories to static, straight and symmetric bifurcations. 

\subsection{Stable fixed point for a symmetric bifurcation}\label{sec:stable_fixed_point}

When two geodesic fingers grow off of the tip of a straight, semi-infinite mother branch, they converge towards an opening angle of 72$^{\circ}$, regardless of initial conditions \cite{selander}. When the bifurcation is symmetric about the mother branch, we can express analytically the path towards the fixed point. In section~\ref{sec:geo_growth}, for instance, we have determined the path of two fingers growing off of the sides of the mother branch (Fig.~\ref{fig:geodesic_growth_two_tips}).

\begin{figure}
\centering
\includegraphics[width = \linewidth]{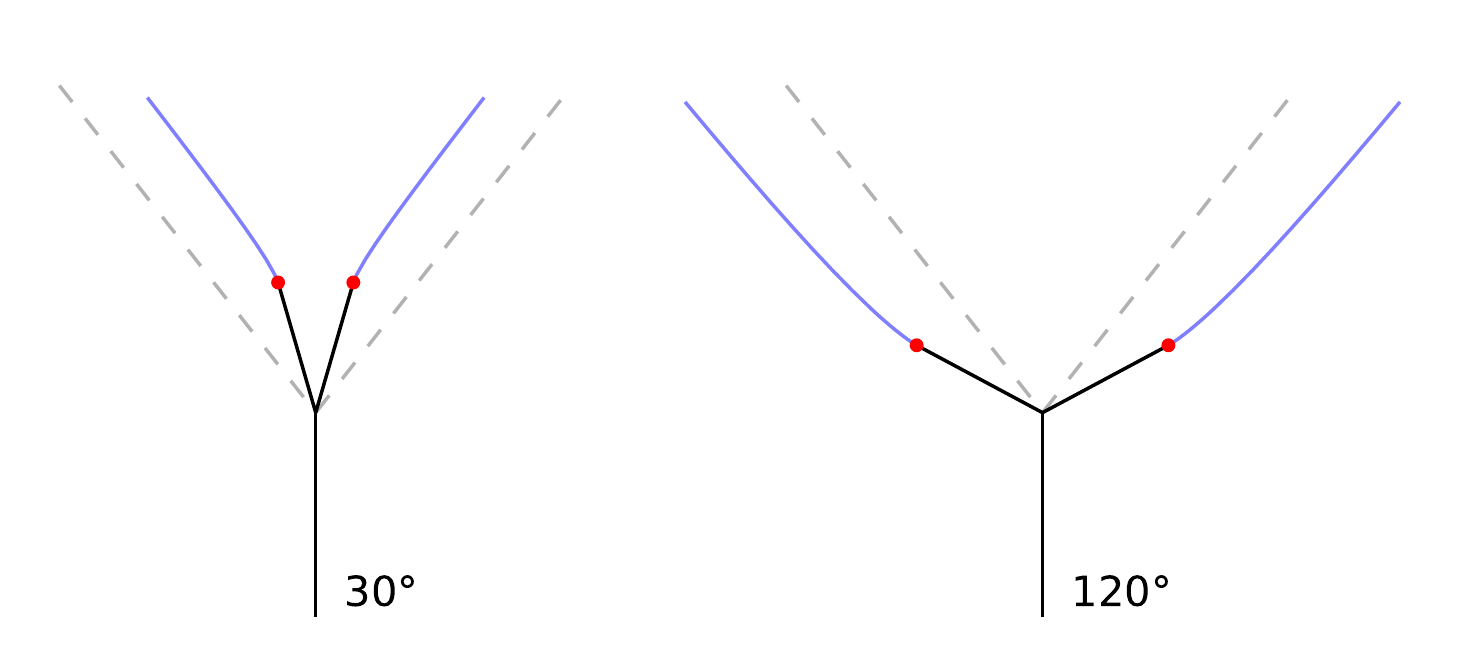}
\caption{ Symmetric fingers (blue lines) growing off of an initially straight bifurcation (black lines). Fingers are geodesic. Dashed lines indicate an angle of 72$^{\circ}$. Red dots show starting points.\label{fig:fish_tail_bif_growth}}
\end{figure}

Here, we follow the same procedure to grow symmetric fingers, starting from a straight bifurcation with daughter branches of length unity (Fig.~\ref{fig:fish_tail_bif_growth}). We use equation~(\ref{eq:integrated_loewner}), which defines the evolution of the mapping $g$ as the pole $a$ moves, starting from the initial mapping $g_0$. Next we substitute $g_c$, the mapping corresponding to a straight bifurcation of angle $2 \pi c$ (equation~(\ref{eq:bifurcation_map})), for the initial condition $g_0$. Finally, we apply the resulting expression to the tip of a finger ($z = \gamma$) and find:
\begin{equation}
4 a^{5/2} = 5 g_c(\gamma)^{1/2} a_c^2 - g_c(\gamma)^{5/2} \, ,
\label{eq:piotr_trajectory}
\end{equation}
where $a_c$ is the initial position of the pole:
\begin{equation}
a_c = \left( \dfrac{1}{c}- 1 \right)^{ (c-1)/2 } \, .
\end{equation}
At any time, we solve equation~(\ref{eq:piotr_trajectory}) for $g_c(\gamma)$ and map the corresponding trajectories with the inverse map $f_c$. Figure~\ref{fig:fish_tail_bif_growth} shows the result of this procedure for two initial bifurcations. As expected, we find that the fingers approach an opening angle of 72$^{\circ}$, regardless of the initial opening angle.

Therefore, geodesic growth, which maximizes dynamically the flux into tips, brings the fingers to a fixed point. This suggests that the fixed point corresponds to an optimal bifurcation. Next we assess this proposition, and propose a graphical interpretation of the fingers' convergence towards the fixed point.

\subsection{Optimal trajectories vs. optimal shapes}\label{sec:optimal_vs}

\begin{figure*}
\centering
\includegraphics[width = 0.85\linewidth]{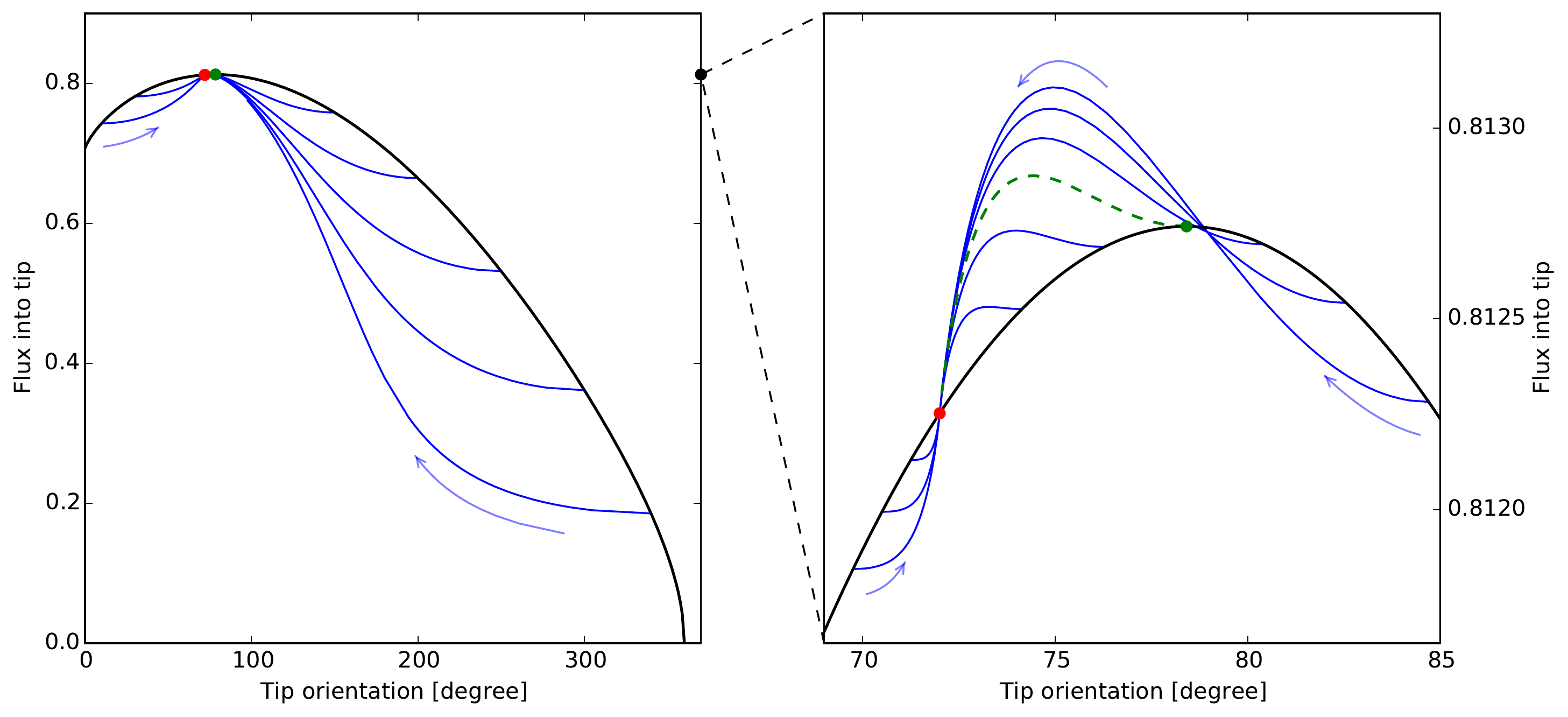}
\caption{Flux into tip as a function of tip orientation. Black line represents straight, symmetric bifurcations (equation~(\ref{eq:straight_bif_c2})). Blue lines are trajectories of geodesic fingers starting from distinct initial bifurcations (equation~(\ref{eq:geo_c2})). Arrows indicate time direction. Dots correspond to the $72^{\circ}$ bifurcation (red, $c = 1/5$), and to the maximum flux (green, $c = c_{\mathrm{max}}$). Dashed green trajectory starts from the maximum flux.}\label{fig:flux}
\end{figure*}

Geodesic fingers co-evolve with the field that surrounds them. Their dynamics are therefore set in an infinite-dimensional phase space. To represent the growth of a symmetric bifurcation, we choose the two-dimensional subspace parameterized by the orientation of the finger's tip, and the intensity of the flux into it. The location of a finger in this subspace depends only on the complex parameter $c_2$, which characterizes the behavior of the mapping $f$ near the tip: $1/\sqrt{c_2}$ is the flux, and the argument of $c_2$ the orientation of the tip (Sec.~\ref{sec:PLS}).

We first calculate the coefficient $c_2$ for a straight, symmetric bifurcation with opening angle $2 \pi c$ and branches of length unity. This static bifurcation will serve as a reference for the trajectories of geodesic fingers in the flux-orientation space. To to so, we differentiate $f_c$ twice at a pole. We find
\begin{equation}
  c_2 = \dfrac{2}{ ( 1 - 1/c )^c } \, .
\label{eq:straight_bif_c2}
\end{equation}
Figure~\ref{fig:flux} shows the curve defined by this equation in the flux-orientation space. As the parameter $c$ varies, this curve reaches a maximum flux for $c =  c_{\mathrm{max}}$, which corresponds to the static bifurcation that optimizes the flux. By definition, its opening angle in the physical space is $2 \pi c_{\mathrm{max}}$. To calculate this angle, we minimize the absolute value of $c_2$ with respect to the parameter $c$, according to Eq.~(\ref{eq:straight_bif_c2}):
\begin{equation}
(c_{\mathrm{max}}-1) \log \left(\frac{c_{\mathrm{max}}}{1-c_{\mathrm{max}}}\right) = 1 \, .
\end{equation}
Numerically, we find $c_{\mathrm{max}} \approx 0.218$, with a corresponding bifurcation angle of about $78.5^{\circ}$, slightly wider than the angle resulting from geodesic growth. This surprising finding shows that the fixed point of geodesic growth does not correspond to the bifurcation that maximizes the flux (red and green dots on figure~\ref{fig:flux}).

To understand how optimal dynamics can lead to a sub-optimal configuration, we now turn our attention to the dynamics of this convergence in the flux-orientation space. We consider the straight, symmetric bifurcation corresponding to $f_c$ as an initial condition, off of which two geodesic fingers grow. To represent their growth in the flux-orientation space, we need to calculate the evolution of the coefficient $c_2$ as they grow. To do so, we return to equation~(\ref{eq:bifurcation_map}) and, like in Section~\ref{sec:stable_fixed_point}, we substitute $g_c$ for $g_0$. This time, however, we need to expand equation~(\ref{eq:bifurcation_map}) in the tip's neighborhood. A straightforward, although cumbersome, calculation leads to
\begin{equation}
c_2 = - \dfrac
{ \sqrt{ a \, g_c(\gamma) } }
{ a_c^2 - g_c(\gamma)^2  }
\,
f'_c \left( g_c(\gamma) \right) \, .
\label{eq:geo_c2}
\end{equation}
Inverting equation~(\ref{eq:piotr_trajectory}) yields $g_c(\gamma)$ as the pole $a$ moves, and the above expression defines the evolution of $c_2$ as the fingers grow. 

Figure~\ref{fig:flux} shows the evolution of $c_2$ in the flux-orientation space. Since all fingers begin their geodesic growth from an initially straight bifurcation, the corresponding trajectories of $c_2$ in the flux-orientation space all start from the curve defined by equation (\ref{eq:straight_bif_c2}). As expected, all trajectories converge towards the $72^{\circ}$ bifurcation (red dot on Fig.~\ref{fig:flux}), which lies on the curve defined by equation (\ref{eq:straight_bif_c2}). In particular, the trajectory starting from the optimal straight bifurcation ($c = c_{\mathrm{max}}$, green dashed line on Fig.~\ref{fig:flux}) also returns to the fixed point ($c = 1/5$).

Most fingers monotonically increase the flux on their way towards the fixed point. In that sense, the optimal growth rule defined in section~\ref{sec:max_flux} is often efficient: as the finger advances, its tip consistently collects more flux. However, a closer look at the neighborhood of the fixed point reveals a small family of more exotic trajectories. They, too, begin with an increase of the flux, but they later reach a maximum before returning to the fixed point. Although they follow an optimal trajectory, they end up decreasing their flux as they grow. Mathematically, this indicates that the right-hand term in equation~(\ref{eq:maxi_4}) changes sign during growth.
The trajectory starting from the optimal straight bifurcation belongs to this family: optimal growth drives it away from the optimal static configuration it started at, to bring it to the fixed point.

Geodesic fingers grow in locally optimal directions consistent with the evolving diffusion field that surrounds them. Somewhat counter to intuition, this does not bring them to a final, optimal shape. This behavior illustrates the distinction between a network that grows according to optimal dynamics, and a static, optimal network \cite{maritan1996universality}. In a nutshell, fingers that optimize their growth at each time to maximize the flux into their tip might not end up in an optimal configuration.

\section{Conclusion}

To represent the growth of a ramified network, we need to specify (at least) three basic rules, which represent respectively the growth velocity, the growth direction, and the nucleation of new branches (tip splitting and side-branching). Such networks often form in response to the diffusive field they drain, which make them amenable the Loewner formalism. Here, after revisiting previous contributions, we have shown how the three rules controlling the growth of a Laplacian network translate straightforwardly into this formalism:
\begin{enumerate}
\item The growth factor $G_n$ sets the velocity of the $n$-th tip;
\item The motion of the pole $a_n$ controls the direction in which the tip grows;
\item The creation of a pole generates a new branch. 
\end{enumerate}

We have then examined three formulations of the second rule: geodesic growth, local symmetry, and maximization of the diffusion flux. A previous contribution showed the equivalence between the two first rules \cite{yossi}. Here, we have established the remaining equivalence, thus relating local symmetry to dynamical optimality. These equivalences suggest that a variety of systems could generate geodesic networks, possibly for distinct physical reasons.

One way to identify a network of this class is to recognize its shape. For instance, river networks form over geological timescales, and we have no direct access to the dynamics of their growth. Among them, a drainage network formed by seepage erosion in the Apalachicola River Basin, Florida shows an average branching angle of $71.9^{\circ} \pm 0.8^{\circ}$, remarkably close to $72^{\circ}$, the expected value for a geodesic network \cite{devauchelle2012ramification}. In this case, field evidence suggests that groundwater plays the role of the diffusive field.

Based on the equivalence between geodesic growth and local optimization, we can interpret the special bifurcation angle of geodesic networks as the geometrical signature of the competition between nascent branches. In the Florida drainage network, each river grows in the direction where it will extract the most groundwater. However, as illustrated in Section~\ref{sec:optimal_vs} of the present contribution, this does not ensure that the present geometry of this network optimizes any physical quantity \cite{arneodo1989uncovering,Rinaldo1992,maritan1996universality,Errera1998,Bejan2006}.

To summarize, the Loewner equation formalizes the translation of local growth dynamics into the geometry of the network they generate. Can we use it backward, to infer the history of a network from its final geometry? Recent steps in this direction suggest that the complete geometry of a drainage network, beyond its branching angle, can indeed inform us about its growth \cite{abrams2009growth,yossi}.

\subsection*{Acknowledgments}
The authors would like to thank A.P.~Petroff, R.S~Yi, E.~Stansifer and J.~Mathiesen for stimulating discussions. O.~D. is also grateful to E.~Lajeunesse and F.~Métivier for their reading of the manuscript and their suggestions.

This work was supported by the National Science Centre, Poland under research grant 2012/07/E/ST3/01734 and the \emph{Émergence(s)} program of the \emph{Marie de Paris}, France. The contributions of O.~D., Y.~C., H.J.~S., and D.H.~R are based in part upon work
supported by the U.S. Department of Energy, Office of Science, Office
of Basic Energy Sciences, Chemical Sciences, Geosciences, and
Biosciences Division under Award Number FG02-99ER15004.

%%%%%%%%%%%%%%%%%%%%%%%%%%%%%%%%%%%%%%%%%%%ù

\bibliographystyle{ieeetr}
\bibliography{biblio_loewner}% Produces the bibliography via BibTeX.

\end{document}